\documentclass[prd,tightenlines,nofootinbib,superscriptaddress]{revtex4}
\usepackage{amsfonts,amsmath,amssymb,amsthm,bbm}

\usepackage{color,psfrag}
\usepackage[dvips]{graphicx}

\newtheorem{res}{Result}
\newtheorem{conj}{Conjecture}

\newcommand{\C}{{\mathbb C}}
\newcommand{\N}{{\mathbb N}}
\newcommand{\R}{{\mathbb R}}
\newcommand{\Z}{{\mathbb Z}}

\newcommand{\cA}{{\mathcal A}}
\newcommand{\cE}{{\mathcal E}}

\newcommand{\cH}{{\mathcal H}}
\newcommand{\dcH}{{{}^2{\mathcal H}}}
\newcommand{\cHi}{{{}^2{\mathcal H}_{inv}}}
\newcommand{\cQ}{{\mathcal Q}}

\newcommand{\SU}{\mathrm{SU}}
\newcommand{\SL}{\mathrm{SL}}

\newcommand{\U}{\mathrm{U}}

\newcommand{\vJ}{\vec{J}}
\newcommand{\vcJ}{\vec{{\cal J}}}

\newcommand{\id}{\mathbb{I}}
\def\om{\omega}
\def\tom{{\widetilde{\omega}}}

\newcommand{\be}{\begin{equation}}
\newcommand{\ee}{\end{equation}}
\newcommand{\beq}{\begin{eqnarray}}
\newcommand{\eeq}{\end{eqnarray}}
\newcommand{\bea}{\begin{eqnarray}}
\newcommand{\eea}{\end{eqnarray}}
\newcommand{\nn}{\nonumber}

\newcommand{\bin} [2] {\left (\begin{array}{c}#2\\#1\end{array} \right ) }

\newcommand{\su}{{\mathfrak su}}
\renewcommand{\sl}{{\mathfrak sl}}

\renewcommand{\u}{{\mathfrak u}}

\newcommand{\la}{\langle}
\newcommand{\ra}{\rangle}

\newcommand{\tr}{{\mathrm Tr}}
\newcommand{\f}{\frac}
\newcommand{\tl}{\widetilde}
\newcommand{\vphi}{\varphi}
\def\eps{\epsilon}

\newcommand{\hh}{{\mathbf h}}
\newcommand\vj{{\vec{{\mathfrak j}}}}

\newcommand{\bz}{\overline{z}}

\newcommand{\Ref}[1]{(\ref{#1})}

\def\nn{\nonumber}

\def\arr{\rightarrow}

\def\tk{\widetilde{k}}

\def\te{\tl{e}}

\def\vphi{\varphi}

\def\Ea{E^{(\alpha)}}
\def\Eb{E^{(\beta)}}
\def\Fa{F^{(\alpha)}}
\def\Fb{F^{(\beta)}}
\def\dag{^\dagger}
\def\bz{{\bf z}}

\begin{document}

\title{Dynamics for a 2-vertex Quantum Gravity Model}

\author{{\bf Enrique F. Borja}}
\affiliation{Institute for Theoretical Physics III, University of
Erlangen-N\"{u}rnberg, Staudtstra{\ss}e 7, D-91058 Erlangen (Germany)}
\affiliation{Departamento de F\'{\i}sica Te\'{o}rica and IFIC, Centro Mixto
Universidad de Valencia-CSIC. Facultad de F\'{\i}sica, Universidad de
Valencia, Burjassot-46100, Valencia (Spain)}
\author{{\bf Jacobo D\'{\i}az-Polo}}
\affiliation{Institute for Gravitation and the Cosmos \& Physics
Department, Penn State, University Park, PA 16802-6300, U.S.A.}
\author{{\bf I\~naki Garay}}
\affiliation{Institute for Theoretical Physics III, University of
Erlangen-N\"{u}rnberg, Staudtstra{\ss}e 7, D-91058 Erlangen (Germany)}
\author{{\bf Etera R. Livine}\email{etera.livine@ens-lyon.fr}}
\affiliation{Laboratoire de Physique, ENS Lyon, CNRS-UMR 5672, 46
All\'ee d'Italie, Lyon 69007, France}

\date{\today}

\begin{abstract}
We use the recently introduced U(N) framework for loop quantum gravity
to study the dynamics of spin network states on the simplest class of
graphs: two vertices linked with an arbitrary number N of edges. Such
graphs represent two regions, in and out, separated by a boundary
surface. We study the algebraic structure of the Hilbert space of spin
networks from the U(N) perspective. In particular, we describe the
algebra of operators acting on that space and discuss their relation
to the standard holonomy operator of loop quantum gravity.
Furthermore, we show that it is possible to make the restriction to
the isotropic/homogeneous sector of the model by imposing the
invariance under a global U(N) symmetry. We then  propose a U(N)
invariant Hamiltonian operator and study the induced dynamics.
Finally, we explore the analogies between this model and  loop quantum
cosmology and sketch some possible generalizations of it.
\end{abstract}

\maketitle


\tableofcontents
\pagebreak

\section*{Introduction}

%
%
%
%
%

Loop quantum gravity (LQG) presents a rigorous framework towards the
quantization of general relativity. The main issues faced by this
quantization scheme are the precise dynamics of the theory and the
derivation of its semi-classical limit. There has been tremendous
progress on these topics since the original formulation of the
theory, but a punch line is still missing. Our main motivation for
the present work is the study of the LQG dynamics. Since Thiemann's
original proposal of a well-defined Hamiltonian constraint operator
for LQG \cite{thomas1,thomas2}, there has been various more recent
proposals among which we point out the algebraic quantum gravity
framework \cite{tina1} and the spinfoam approach e.g.
\cite{spinfoam1,spinfoam2}. Our main source of inspiration for our
current approach is the recent model introduced by Rovelli and
Vidotto \cite{carlo1}. The logic behind their model is to implement
the LQG dynamics on the simplest non-trivial class of spin network
states, thus constructing a first order truncation of the full
theory. They considered spin network states based on a fixed graph
with two vertices related by four edges, so that their model can be
called ``tetrahedron LQG model". Very interestingly, it was shown
that this model can be understood as reproducing a cosmological
setting in LQG and leads to a physical framework very similar to
loop quantum cosmology \cite{carlo1,carlo2}. It was also shown that
the same procedure can be successfully applied to the current
spinfoam models \cite{carlo3}. This ``dipole quantum cosmology" is
the starting point of our work.

We consider the generalization of the Rovelli-Vidotto model to spin network states based on a graph with still 2 vertices but now with an arbitrary number $N$ of edges. From their viewpoint, this should allow to introduce more anisotropy/inhomogeneity in their model. Here, we start anew with a thorough study of the algebraic structure of the Hilbert space of spin  networks on the 2-vertex graph. To this aim, we use the recently developed $\U(N)$ framework for $\SU(2)$-intertwiners \cite{un1,un2,un3}. Considering our spin networks as two intertwiners (one for each vertex) glued by matching conditions representing the $N$ edges between the two vertices, we use the operators of the $\U(N)$ formalism to act on our Hilbert space. From there, we introduce two improvements on this standard setting. First, we give the explicit link between the $\U(N)$ operators and the usual holonomy operators of LQG, which provides a dictionary to translate between the standard approach to LQG and the more recent $\U(N)$ framework. Second, we identify a $\U(N)$ symmetry generated by operators acting on the coupled system of the two vertices and which reduces the full space of arbitrary spin network states to the reduced space of homogeneous/isotropic states.

Then we finally introduce $\U(N)$-invariant operators, which respect the isotropy/homogeneity of the quantum states, and use them to build a $\U(N)$-invariant Hamiltonian operator encoding the dynamics of our 2-vertex model. We study the algebra of our operators and propose an interpretation of our creation operator as a ``black hole" creation operator. Finally, we study the spectral properties of our Hamiltonian and discuss in detail the interplay between our framework and loop quantum cosmology. We conclude with the limitations of our model and its possible generalizations in order to take into account more physical situations.

\section{The $\U(N)$ Framework for LQG Intertwiners}


Let us focus on the intertwiner spaces for the Lie group $\SU(2)$
and define the Hilbert space of intertwiners between $N$
irreducible representations of spin $j_1,..,j_N$~:
\be
\cH_{j_1,..,j_N}\,\equiv\, \textrm{Inv}[V^{j_1}\otimes..\otimes V^{j_N}].
\ee
The object that will be most relevant to us is actually the space of
intertwiners with $N$ legs and fixed total area $J=\sum_i j_i$~:
\be
\cH_N^{(J)}\,\equiv\,\bigoplus_{\sum_i j_i=J}\cH_{j_1,..,j_N}.
\ee
We also introduce the full Hilbert space of $N$-valent intertwiners:
\be
\cH_N\,\equiv\,\bigoplus_{\{j_i\}}\cH_{j_1,..,j_N}\,=\, \bigoplus_{J\in\N}\cH_N^{(J)}.
\ee

The key starting point to our dynamical model is that there is a
natural action of $\U(N)$ on the intertwiner space $\cH_N$
\cite{un1}, that more precisely the intertwiner spaces $\cH_N^{(J)}$
carry irreducible representations of $\U(N)$ \cite{un2} and finally
that the full space $\cH_N$ can be endowed with a Fock space
structure with creation and annihilation operators compatible with
the $\U(N)$ action \cite{un3}.

This $\U(N)$ formalism is based on the Schwinger representation of
the $\su(2)$ Lie algebra in terms of harmonic oscillators. Let us
introduce $2N$ oscillators with creation operators $a_i,b_i$ with
$i$ running from 1 to $N$:
$$
[a_i,a\dag_j]=[b_i,b\dag_j]=\delta_{ij}\,,\qquad [a_i,b_j]=0.
$$
The generators of the $\SU(2)$ transformations acting on each leg of
the intertwiner are realized as quadratic operators in terms of the
oscillators:
\be
J^z_i=\f12(a\dag_i a_i-b\dag_ib_i),\qquad
J^+_i=a\dag_i b_i,\qquad
J^-_i=a_i b\dag_i,\qquad
E_i=(a\dag_i a_i+b\dag_ib_i).
\ee
The $J_i$'s satisfy the standard commutation algebra while the total
energy $E_i$  is a Casimir operator:
\be
[J^z_i,J^\pm_i]=\pm J^\pm_i,\qquad
[J^+_i,J^-_i]=2J^z_i,\qquad
[E_i,\vec{J}_i]=0.
\ee
The correspondence with the standard $|j,m\ra$ basis of $\su(2)$
representations is simple.  The eigenvalue $m_i$ of $J^z_i$ is given by
the half-difference of the energies between the two oscillators,
while the total energy $E_i$ gives twice the spin, $2j_i$, living on
the $i$-th leg of the intertwiner:
\be
|n_a,n_b\ra_{OH}=|\f12(n_a+n_b),\f12(n_a-n_b)\ra\,,\qquad
|j,m\ra=|j+m,j-m\ra_{OH}
\ee

Intertwiner states are by definition invariant under the global
$\SU(2)$ action, generated by:
\be
J^z=\sum_{i=1}^N J^z_i,\qquad
J^\pm=\sum_i J^\pm_i.
\ee
Then, operators acting on the intertwiner space need to commute with
these operators too. The simplest family of invariant operators was
identified in \cite{un1} and are quadratic operators acting on
couples of legs:
\be
E_{ij}=a\dag_ia_j+b\dag_ib_j, \qquad
E_{ij}\dag=E_{ji}.
\ee
The starting result is that these operators are invariant under
global $\SU(2)$ transformations and form a $\u(N)$ algebra:
\be
[\vec{J},E_{ij}]=0,\qquad
[E_{ij},E_{kl}]\,=\,
\delta_{jk}E_{il}-\delta_{il}E_{kj}.
\ee
The diagonal operators $E_i\equiv E_{ii}$ form the Cartan
sub-algebra of $\u(N)$, while the off-diagonal operators $E_{ij}$
with $i\ne j$ are the raising and lowering operators. As said
earlier, the generators $E_i$ give twice the spin $2j_i$ while the
$\U(1)$ Casimir $E=\sum_i E_i$ will give twice the same area,
$2J\equiv \sum_i 2j_i$. Then all operators $E_{ij}$ commute with the
$\U(1)$ Casimir, thus leaving the total area $J$ invariant:
\be
[E_{ij},E]=0.
\ee

The usual $\SU(2)$ Casimir operators have a simple expression in
terms of these $\u(N)$ generators\footnotemark:
\be
(\vec{J}_i)^2=\f{1}{2}E_i\left(\f{E_i}{2}+1\right),\qquad
\forall i\ne j,\,\,
(\vec{J}_i\cdot \vec{J}_j)
\,=\,
\f12E_{ij} E_{ji} -\f14E_iE_j-\f12 E_i.
\label{scalarop}
\ee
\footnotetext{ Let us point out that the case $i=j$ of
$(\vec{J}_i\cdot \vec{J}_j)$ does not give back exactly the formula
for $(\vec{J}_i)^2$ due to the ordering of the oscillator operators.
The two formula agree on the leading order quadratic in $E_i$ but
disagree on the correction linear in $E_i$. }
Moreover, the explicit definition of the $E_{ij}$'s in terms of
harmonic oscillators leads to quadratic constraints on these
operators as shown in \cite{un2}. Here, we give a slight
generalization of these constraints:
\be
\forall i,j,\quad
\sum_k E_{ik}E_{kj}=E_{ij} \left(\f E2+N-2\right),
\ee
where we have assumed that the global $\SU(2)$ generators $\vec{J}$
vanish.
This can be interpreted as the square of the matrix $E_{ij}$ giving
back itself up to a factor which is its trace $E$ divided by 2 up to
an ordering factor $(N-2)$. Classically, considering a (Hermitian)
matrix $M$ satisfying the similar constraint $M^2=\f12(\tr \,M)M$,
it is straightforward to show that $M$ has only a single eigenvalue
with degeneracy 2. We will see below that this interpretation
actually holds at the quantum level. This viewpoint of seeing the
$\u(N)$ framework for intertwiners as the quantization of a
classical matrix model will be investigated in detail elsewhere
\cite{matrixmodel}. These quadratic constraints on the $E_{ij}$
operators lead to non-trivial restrictions on the representations of
$\u(N)$ obtained from our construction. To solve them, the method
used in \cite{un2} is to apply them to a highest weight vector.
Thus let us introduce a highest weight vector $v$, such that it
diagonalizes the generators of the Cartan sub-algebra $E_i\,v=
l_i\,v$ with the eigenvalues $l_1\ge l_2 \ge l_3\ge..\ge 0$ and that
it vanishes under the action of the raising operators $E_{ij}\,v=0$
for all $i<j$. Applying the quadratic constraints given above to
such vector $v$, one shows that the eigenvalues must necessarily be
of the following form, $[l_1,l_2,l_3..,l_N]=[l,l,0,..,0]$.

Coming back to the intertwiner spaces, we get that the intertwiner
space $\cH_N^{(J)}$ at fixed total area $J=\sum_i j_i$ carries an
irreducible representation of $\U(N)$ with highest weight
$[J,J,0,..,0]$ with $l=J$, as was shown in \cite{un2}. This highest
weight vector corresponds to a bivalent intertwiner with $j_1=j_2=\f
J2$ and $j_3=..=j_N=0$. Then one acts on it with $\u(N)$ operators
to get all possible intertwiner states with the same total area $J$.
Finally, the value of the quadratic $\U(N)$ Casimir is easily
obtained from the previous constraints:
\be
\sum_{i,k} E_{ik}E_{ki}=E\left(\f E2+N-2\right)= 2J(J+N-2).
\label{casimir}
\ee
And we can also compute the dimension of $\cH_N^{(J)}$ in terms of
binomial coefficients using the standard formula for $\U(N)$
representations:
\be
D_{N,J}\,\equiv\,
\dim \cH^{(J)}_N
\,=\,
\f{1}{J+1}\bin{J}{N+J-1}\bin{J}{N+J-2}
\,=\,
\frac{(N+J-1)!(N+J-2)!}{J!(J+1)! (N-1)!(N-2)!}\,.
\label{dimNJ}
\ee

\medskip

Next, we introduce annihilation and creation operators to move
between the spaces $\cH_N^{(J)}$ with different areas $J$ within the
bigger Hilbert space of all intertwiners with $N$ legs \cite{un3}.
We define the new operators:
\be
F_{ij}=(a_i b_j - a_j b_i),\qquad
F_{ji}=-F_{ij}.
\ee
These are still invariant under global $\SU(2)$ transformations, but
they do not commute anymore with the total area operator $E$. They nevertheless form a closed algebra together with the operators $E_{ij}$:
\bea
{[}E_{ij},E_{kl}]&=&
\delta_{jk}E_{il}-\delta_{il}E_{kj}\nn\\
{[}E_{ij},F_{kl}] &=& \delta_{il}F_{jk}-\delta_{ik}F_{jl},\qquad
{[}E_{ij},F_{kl}^{\dagger}] = \delta_{jk}F_{il}^{\dagger}-\delta_{jl}F_{ik}^{\dagger}, \\
{[} F_{ij},F^{\dagger}_{kl}] &=& \delta_{ik}E_{lj}-\delta_{il}E_{kj} -\delta_{jk}E_{li}+\delta_{jl}E_{ki}
+2(\delta_{ik}\delta_{jl}-\delta_{il}\delta_{jk}), \nn\\
{[} F_{ij},F_{kl}] &=& 0,\qquad {[} F_{ij}^{\dagger},F_{kl}^{\dagger}] =0.\nn
\eea
The annihilation operators $F_{ij}$ allow to go from $\cH_N^{(J)}$ to
$\cH_N^{(J-1)}$ while the creation operators $F\dag_{ij}$ raise the
area and go from $\cH_N^{(J)}$ to $\cH_N^{(J+1)}$. In \cite{un3},
these operators were mainly used in order to construct coherent
states transforming consistently under $\U(N)$ transformations.
Here, we will focus on their use as building blocks for the dynamics
of intertwiners.

These new operators satisfy new quadratic constraints together with
the $E_{ij}$.  Starting with their explicit definitions in  terms of
oscillators, it is easy to get on the intertwiner space (assuming
global $\SU(2)$ invariance, $\vec{J}=0$):
\beq
&&\sum_k F^\dagger_{ik}E_{jk} = F^\dagger_{ij}\, \frac{E}{2}, \qquad\qquad\quad
\sum_k E_{jk} F^\dagger_{ik} = F^\dagger_{ij}\left(\frac{E}{2}+N-1\right),\label{constraint1}\\
&&\sum_k E_{kj}F_{ik} = F_{ij}\, \left(\frac{E}{2}-1\right), \qquad
\sum_k F_{ik} E_{kj}  = F_{ij}\left(\frac{E}{2}+N-2\right),\label{constraint2}\\
&&\sum_k F^\dagger_{ik}F_{jk} = E_{ij}
\left(\frac{E}{2}+1\right),\qquad
\sum_k F_{jk}F\dag_{ik} = (E_{ij}+2\delta_{ij})
\left(\frac{E}{2}+N-1\right)\,.\label{constraint3}
\eeq
These relations will be useful later for the analysis of the
dynamics of our 2-vertex toy model. It is also interesting  to look at them from a classical matrix perspective. Indeed ignoring the ``quantum" ordering terms, we consider the classical Hermitian matrix $M=M\dag$ corresponding to the operators $E_{ij}$ and the classical antisymmetric matrix $Q=-Q^t$ corresponding to the operators $F_{ij}$ and the quadratic constraints become simply:
\beq
&&M^2=\lambda\,M,\qquad \bar{Q}Q=-\lambda \,M, \nn\\
&&QM=\lambda\,Q,\qquad M\bar{Q}=\lambda\,\bar{Q},
\eeq
where the eigenvalue is defined in terms of $M$ as $\lambda=(\tr\,
M)/2$. These conditions correspond exactly to the relations between
the matrices $\rho$ and ${\bf z}$ introduced in \cite{un3} in order
to define $\U(N)$ coherent states. This strongly suggests that this
$\U(N)$ framework can be derived as the quantization of a matrix
model. This will be investigated in detail in future work
\cite{matrixmodel}.

\section{The 2-Vertex Graph and the $\U(N)$ Symmetry}

\subsection{The 2-Vertex Algebraic Structure}

\begin{figure}[h]
\begin{center}
\includegraphics[height=30mm]{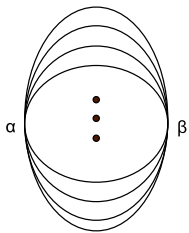}
\caption{The 2-vertex graph: the two vertices $\alpha$ and $\beta$
are linked by $N$ edges. \label{2vertex}}
\end{center}
\end{figure}

Let us consider the simplest non-trivial graph for spin network
states in Loop Quantum Gravity: a graph with two vertices linked by
$N$ edges, as shown in fig.\ref{2vertex}. This is a generalization
of the simplest tetrahedral model introduced by Rovelli and Vidotto
in \cite{carlo1} and which was shown to be related to models of
quantum cosmology \cite{carlo2}. Their model is made of two vertices
related by exactly four edges, which can be interpreted as two
tetrahedra glued together. Here we consider the more general case of
an arbitrary number $N$ of edges.

Naming the two vertices $\alpha$ and $\beta$, we now have two
intertwiner spaces attached respectively to $\alpha$ and $\beta$
with their own operators $\Ea_{ij},\Fa_{ij}$ and
$\Eb_{ij},\Fb_{ij}$. The total Hilbert space of these uncoupled intertwiners is the tensor product of the two spaces of $N$-valent intertwiners:
\be
\cH_{\otimes 2}
\,=\,
\cH_N\otimes\cH_N
=\bigoplus_{J_\alpha,J_\beta}\cH_N^{(J_\alpha)}\otimes\cH_N^{(J_\beta)}
=\bigoplus_{\{j_i^\alpha,j_i^\beta\}} \cH_{j_1^\alpha,..,j_N^\alpha}\otimes\cH_{j_1^\beta,..,j_N^\beta}.
\ee
This space carries two decoupled $\U(N)$-actions, one acting on the intertwiner space associated to the vertex $\alpha$ and the other acting on $\beta$.

However, considering the 2-vertex graph, the two intertwiner spaces are not
independent. There are matching conditions \cite{un2} imposing that each edge
carries a unique $\SU(2)$ representation, thus the spin on that edge must be the same
seen from $\alpha$ or $\beta$ i.e. $j_i^\alpha=j_i^\beta$. This translates into the fact that the
oscillator energy for $\alpha$ on the leg $i$ must be equal to the energy for $\beta$ on its $i$-th leg:
\be
\cE_i\,\equiv\,\Ea_i -\Eb_i
\,=\,0.
\ee
This set of conditions is stronger than requiring that the total
area of $\alpha$ is the same as $\beta$, even though this is a
necessary condition. Then, the Hilbert space of spin network states
on the 2-vertex graph is much smaller than the decoupled Hilbert space $\cH_{\otimes 2}$:
\be
\dcH\equiv\bigoplus_{\{j_i\}}\cH_{j_1,..,j_N}^{(\alpha)}\otimes \cH_{j_1,..,j_N}^{(\beta)}.
\ee
Then, in order to define consistent operators acting on $\dcH$, we
not only have to check that they are invariant under global $\SU(2)$
transformations, but we also have to check that they commute
(at least weakly) with the matching conditions $\cE_i$.


We can construct operators deforming consistently the boundary
between $\alpha$ and $\beta$. We introduce the following symmetric operators:
\be
e_{ij}\equiv\Ea_{ij}\Eb_{ij},\qquad
f_{ij}\equiv\Fa_{ij}\Fb_{ij},\qquad
f\dag_{ij}\equiv\Fa_{ij}{}\dag\Fb_{ij}{}\dag.
\ee
We check that they commute with the matching conditions
\be
\forall i,j,k,\quad [\cE_k,e_{ij}] \,=\,[\cE_k,f_{ij}] \,=\, 0.
\ee
Calling $E=\sum_i \Ea_i=\sum_i \Eb_i$ the operator giving (twice) the total boundary area on our Hilbert space $\dcH$ satisfying the matching conditions, the operators $e_{ij}$'s preserve the boundary area while the $f_{ij}$'s will modify it:
\be
[E,e_{ij}]=0,\qquad [E,f_{ij}]=-2f_{ij},\qquad [E,f\dag_{ij}]=+2f\dag_{ij}.
\ee
More precisely, the operator $e_{ij}$ increases the spin on the $i$-th edge by
$+\f12$ and decreases the spin of the $j$-th edge. The operator
$f_{ij}$ decreases both spins on the edges $i$ and $j$, while its
adjoint $f\dag_{ij}$ increases both spins by $\f12$. These operators
generate the deformations of the boundary surface, consistently
with both the $\SU(2)$ gauge invariance and the matching conditions
imposed by the graph combinatorial structure. They will be the
building blocks for the dynamics of spin network states on the
2-vertex graph.

Nevertheless these operators $e_{ij},f_{ij},f\dag_{ij}$ do not have simple commutators with each other. Moreover, they are not enough to generate the whole Hilbert space $\dcH$ of spin network states from the vacuum state $|0\ra$. Indeed, they are symmetric in $\alpha\leftrightarrow \beta$ and will only generate states symmetric under the exchange of the two vertices. To explore the whole Hilbert space $\dcH$ and generate the whole space of gauge invariant operators on the 2-vertex graph, we also need operators that act on a single vertex and change states without affecting neither the representations on the edges, nor the intertwiner state living at the other vertex. Natural candidates for such operators acting on $\cH_{j_1,..,j_N}^{(\alpha)}$ are the $\Ea_{ij}\Ea_{ji}$, which are directly related to the usual scalar product operators $\vec{J}^{(\alpha)}_i\cdot\vec{J}^{(\alpha)}_j$ by \Ref{scalarop}. They change the intertwiner living at the vertex $\alpha$ without changing the intertwiner living at $\beta$.
Actually, it is straightforward to see that combining such local operators $\Ea_i,\Ea_{ij}\Ea_{ji}$ and $\Eb_i,\Eb_{ij}\Eb_{ji}$ with the coupled symmetric operators $e_{ij},f_{ij},f\dag_{ij}$ allow to go between any two states in the Hilbert space $\dcH$ of spin network states on our 2-vertex graph and thus generate all gauge invariant operators (i.e the algebra of holonomy  and grasping operators) on that graph.
%

\subsection{The $\U(N)$ Symmetry, the Invariant Subspace and Homogeneous States}


As we have said above, the matching conditions $\cE_k$ break the $\U(N)$-actions on both vertices $\alpha$ and $\beta$. Nevertheless, we can see that the $\cE_k$ generate a $\U(1)^N$ symmetry and that they are part of a larger $\U(N)$ symmetry algebra. Indeed, we introduce the operators:
\be
\cE_{ij}\,\equiv\,
\Ea_{ij}-\Eb_{ji}
\,=\,\Ea_{ij}-(\Eb_{ij})\dag.
\ee
It is straightforward to compute their commutation relations and
check that these operators form a $\u(N)$ algebra:
\be
[\cE_{ij},\cE_{kl}]\,=\,
\delta_{jk}\cE_{il}-\delta_{il}\cE_{kj}.
\ee
The diagonal operators are exactly the matching conditions
$\cE_{kk}=\cE_k$ and generate the Cartan abelian sub-algebra of $\u(N)$.

These $\cE_{ij}$'s generate $\U(N)$ transformations\footnotemark
\,on the  two intertwiner system that act in $\cH_{\otimes 2}$ as
$(U,\bar{U})$ with the transformation on $\beta$ being the complex
conjugate of the transformation on $\alpha$.
\footnotetext{There is another set of $\U(N)$ transformations acting on the
coupled system as $(U,U)$ and generated by:
$$
E^{(+)}_{ij}\,\equiv\,
\Ea_{ij}+\Eb_{ij}.
$$
They satisfy the same commutation relations,
$[E^{(+)}_{ij},E^{(+)}_{kl}]\,=\,\delta_{jk}E^{(+)}_{il}-\delta_{il}E^{(+)}_{kj}$, but they
will not be relevant for our construction.}
In the following, unless otherwise said, we will refer to the
transformations generated by the $\cE_{ij}$ simply as the $\U(N)$ action
for our 2-vertex setting.

Looking closer to the interplay between the matching condition and these new $\u(N)$ generators,
\be
[\cE_{ij},\cE_k]
\,=\,
\delta_{jk}\cE_{ik}-\delta_{ik}\cE_{kj}
\,=\,
(\delta_{jk}-\delta_{ik})\cE_{ij},
\ee
two points are obvious:
\begin{itemize}

\item The operators $\cE_{ij}$ are not fully compatible with the matching conditions and they do not act on the 2-vertex Hilbert space $\dcH$. Thus they do not generate a non-trivial $\U(N)$-action on $\dcH$. Indeed, applying the operator $\cE_{ij}$ on a vector satisfying the matching conditions, $\cE_k\,|\psi\ra=0$, we obtain a vector which does not satisfy the matching conditions unless it is trivial:
    $$
    \cE_k\cE_{ij}\,|\psi\ra
    =[\cE_k,\cE_{ij}]\,|\psi\ra
    =(\delta_{ik}-\delta_{jk})\cE_{ij}\,|\psi\ra.
    $$

\item We can nevertheless look for vectors in  $\dcH$ which  are invariant  under this $\U(N)$ action, $\cE_{ij}\,|\psi\ra=0$ for all $i,j$. In particular, they will satisfy the matching conditions (given by the case $i=j$).

\end{itemize}
Following this line of thought, we introduce the subspace of spin network states which are invariant under this $\U(N)$-action:
\be
\cHi\,\equiv\,
Inv_{\U(N)}\left[\dcH\right]
\,=\,
Inv_{\U(N)}\left[\cH_{\otimes 2}\right]
\,=\,
Inv_{\U(N)}\left[\bigoplus_{J_\alpha,J_\beta}\cH^{(J_\alpha)}_N\otimes\cH^{(J_\beta)}_N\right].
\ee
The first equality holds because the matching conditions $\cE_k$ are part of the $\u(N)$-algebra, while the second equality is simply the definition of the full decoupled Hilbert space $\cH_{\otimes 2}$.
Now taking into account that the spaces $\cH^{(J)}_N$ are irreducible $\U(N)$-representations, requiring $\U(N)$-invariance imposes that the two representations for the two vertices are the same, $J_\alpha=J_\beta$, but furthermore there exists a unique invariant vector in the tensor product $\cH^{(J)}_N\otimes\cH^{(J)}_N$.
%
%
We will call this unique invariant vector $|J\ra$ and we will
construct it explicitly in terms of $e_{ij}$ and $f_{ij}$ in the
section \ref{UNstate} below.

What's important here is that, by imposing $\U(N)$-invariance on our 2-vertex system, we obtain a single state $|J\ra$ for every total boundary area $J$:
\be
\cHi\,=\,\bigoplus_{J\in\N} \C\,|J\ra.
\ee
This means that we have restricted our system to isotropic states, which are not sensitive to area-preserving deformations of the boundary between $\alpha$ and $\beta$. Let us point out that these $\U(N)$-invariant states are both isotropic and homogeneous. Indeed, they are isotropic since all directions are equivalent and that the state only depends on the total boundary area, and they are homogeneous because the quantum state is the same at both vertices $\alpha$ and $\beta$ of the graph.
This has two interesting consequences:
\begin{itemize}

\item It allows to realize the reduction to the isotropic/homogeneous subspace used in the tetrahedral model by Rovelli and Vidotto \cite{carlo1,carlo2} to separate the homogeneous dynamics of the global boundary area $J$ from the inhomogeneous degrees of freedom corresponding to the area-preserving variations of the individual spins $j_i$ living on its edge. Here, we've shown that the projection on isotropic/homogeneous states is achieved by a straightforward $\U(N)$-group averaging.

\item This logic can be applied more generally to loop quantum cosmology, which is based on a symmetry reduction at the classical level  and a quantization {\it \`a la} loop of this reduced phase space. A challenge is to derive loop quantum cosmology from the more general realm of loop quantum gravity. A first important step in this direction was achieved by the tetrahedral model of Rovelli and collaborators \cite{carlo1,carlo2, carlo3} where they could derive a simple isotropic and homogeneous quantum cosmology from loop quantum gravity on the double-tetrahedron graph i.e the 2-vertex graph with $N=4$ edges. Here, we would like to point out that the present $\U(N)$ symmetry allows to perform the symmetry reduction to homogeneous (isotropic) states directly at the quantum level. It seems that this has been an important missing ingredient in order to understand the symmetry reduction from loop quantum gravity to loop quantum cosmology.

\end{itemize}

In the following sections, we will impose this $\U(N)$-invariance, work on the invariant subspace of homogeneous states and look for $\U(N)$-invariant dynamics which would act on this invariant space. In particular, such dynamics will automatically be consistent with the matching conditions.

\subsection{Back to Holonomy Operators}\label{holonomy_section}

In order to understand the precise relation between the new $\U(N)$ formalism that we use here and the more standard framework of loop quantum gravity, it is interesting to investigate the link between our operators $e_{ij}$ and $f_{ij}$ and the usual holonomy and derivation operators of loop quantum gravity.

To define the action of the holonomy operator, let us quickly review the definition of the spin network functionals. We consider the 2-vertex graph and choose to orient all the edges in the same way from the vertex $\alpha$ to the vertex $\beta$. The wave-functions are functions of $N$ group elements in $\SU(2)$ which satisfy a $\SU(2)$-invariance at both vertices:
$$
\psi(g_1,..,g_N)=\psi(h_\alpha^{-1}g_1h_\beta,..,h_\alpha^{-1}g_Nh_\beta),\quad\forall h_\alpha,h_\beta\in\SU(2)^{\times 2}.
$$
Then the (simplest) holonomy operator acts on a couple of edges $(ij)$ by multiplication:
\be
(\chi^{(ij)}\,\psi)(g_1,..,g_N)
\,=\,
\chi_{\f12}(g_ig_j^{-1})\,\psi(g_1,..,g_N),
\ee
where $\chi_{\f12}$ is the spin-$\f12$ character, which takes the trace of the $\SU(2)$ group element in the fundamental representation of spin $\f12$. This operator is clearly Hermitian and symmetric under the exchange $i\leftrightarrow j$.

To understand the action of the holonomy operator, we need the
Clebsh-Gordan coefficients corresponding to tensoring an arbitrary
spin $j$ with the spin $\f12$.  Let us then consider the tensor
product $V^j\otimes V^{\f12}$ which decomposes as is well-known as
$V^{j+\f12}\oplus V^{j-\f12}$. Let us write $\vJ$ and $\vj$ for the
$\su(2)$ generators acting respectively on $V^j$ and $V^{\f12}$
while we call $\vcJ=\vJ+\vj$ the generators of the $\su(2)$
transformations acting on the coupled system. The scalar product
operator $\vJ\cdot\vj$ has two eigenvalues, $\f j2$ and $-\f
{(j+1)}2$, which correspond to the two proper subspaces
$V^{j+\f12}$ and $V^{j-\f12}$. We use the oscillator basis where we
drop the index $OH$,
$$
|n,2j-n\ra\otimes|\eps,1-\eps\ra,
$$
where $n$ running from 0 to $2j$ is the energy of the $a$-oscillator in $V^j$ while $\eps=0,1$ labels the basis of $V^{\f12}$. We remind that the spin $j$ is given by the half-sum of the energies of the two oscillators and the magnetic momentum $m$ is given by half-difference of these energies, so that this state corresponds to $|j,m=n-j\ra\otimes|\f12,\eps-\f12\ra$ in the standard $\SU(2)$ basis.
We can compute explicitly the action of the scalar product operator in this oscillator basis:
\beq
\vJ\cdot\vj\,|n,2j-n\ra\otimes|1,0\ra
&=&
\f12(n-j)|n,2j-n\ra\otimes|1,0\ra
+\f12\sqrt{(n+1)(2j-n)}\, |n+1,2j-n-1\ra\otimes|0,1\ra,\\
\vJ\cdot\vj\,|n,2j-n\ra\otimes|0,1\ra
&=&
\f12(j-n)|n,2j-n\ra\otimes|0,1\ra
+\f12\sqrt{n(2j-n+1)}\, |n-1,2j-n+1\ra\otimes|1,0\ra.
\eeq
From there, it is easy to check that the subspace $V^{j+\f12}$ is spanned by the following basis vectors which are eigenvectors of $\vJ\cdot\vj$ for the eigenvalue $+\f j2$~:
\be
|n,2j+1-n\ra
\,=\,
\f1{\sqrt{2j+1}}
\left(
\sqrt{n}\,|n-1,2j-n+1\ra\otimes|1,0\ra
+\sqrt{2j+1-n}\,|n,2j-n\ra\otimes|0,1\ra
\right),
\ee
for which the eigenvalue of $J^z$ is $m=(n-j)-\f12$. Here the energy level $n$ runs from 0 to $2j+1$.
On the other hand, the basis of the subspace $V^{j-\f12}$ is given by the states orthogonal to the previous ones:
\be
|n-1,2j-n\ra
\,=\,
\f1{\sqrt{2j+1}}
\left(
\sqrt{2j+1-n}\,|n-1,2j-n+1\ra\otimes|1,0\ra
-\sqrt{n}\,|n,2j-n\ra\otimes|0,1\ra
\right),
\ee
for which the eigenvalue of $J^z$ is still $m=(n-j)-\f12$ but now the energy level $n$ runs from 1 to $2j$. We also give the reverse mapping:
\beq
|n,2j-n\ra\otimes|0,1\ra
&=&
\f1{\sqrt{2j+1}}
\left(
\sqrt{2j+1-n}\,|n,2j+1-n\ra
-\sqrt{n}\,|n-1,2j-n\ra
\right), \\
|n,2j-n\ra\otimes|1,0\ra
&=&
\f1{\sqrt{2j+1}}
\left(
\sqrt{n+1}\,|n+1,2j-n\ra
+\sqrt{2j-n}\,|n,2j-1-n\ra
\right).
\eeq
To express the holonomy operator in terms of the $\u(N)$ operators,
it is more convenient to express the previous decomposition formula
in terms of the oscillator operators:
\beq
|n,2j-n\ra\otimes|0,1\ra
&=&
\f1{\sqrt{2j+1}}
\left(
b\dag-a
\right)\,|n,2j-n\ra,
\label{CB1}\\
|n,2j-n\ra\otimes|1,0\ra
&=&
\f1{\sqrt{2j+1}}
\left(
a\dag+b
\right)\,|n,2j-n\ra.
\label{CB2}
\eeq

\medskip

\begin{figure}[h]
\psfrag{a}{$\alpha$}
\psfrag{b}{$\beta$}
\psfrag{d}{$\f12$}
\psfrag{j1}{$j_i$}
\psfrag{j2}{$j_j$}
\psfrag{j1m}{$j_i\pm\f12$}
\psfrag{j2m}{$j_j\pm\f12$}
\begin{center}
\begin{minipage}{3cm}
\includegraphics[height=30mm]{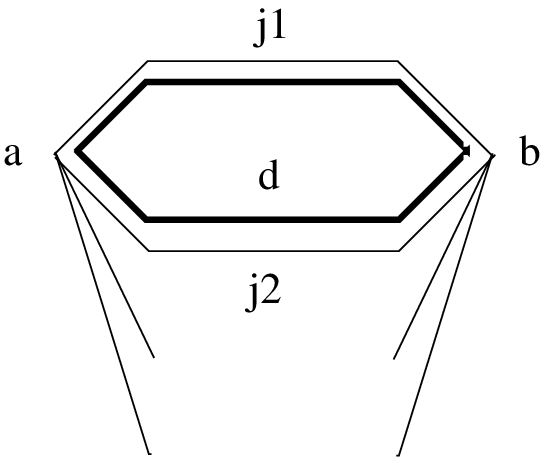}
\end{minipage}
\hspace{5mm}
\begin{minipage}{2cm}
\includegraphics[width=15mm]{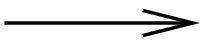}
\end{minipage}
\begin{minipage}{3cm}
\includegraphics[height=30mm]{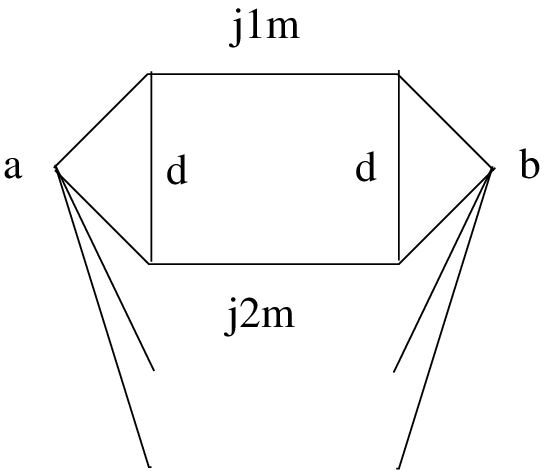}
\end{minipage}
\caption{Holonomy operator $\chi^{ij}$ acting on the couple of edges $(ij)$, which sends the spins $(j_i,j_j)$ to the four possibilities $(j_i+\eta_i,j_j+\eta_j)$ with $\eta_i,\eta_j=\pm \f12$. To study the precise action of this holonomy operator, we focus on its action $\chi^{ij}_\alpha$ and $\chi^{ij}_\beta$ on the individual vertices $\alpha$ and $\beta$. Then we glue back the two resulting intertwiners while respecting the matching conditions.
\label{holonomyfig}}
\end{center}
\end{figure}

Now coming back to the holonomy operator, it can be seen to act independently on each intertwiner at the vertices $\alpha$ and $\beta$, as shown on fig.\ref{holonomyfig}.
%
%
Focusing on the state at the vertex $\alpha$, the holonomy operator
acts on the legs $i$ and $j$ of the intertwiners by tensoring the
state on these legs with the singlet state
$(|1,0\ra\otimes|0,1\ra-|0,1\ra\otimes|1,0\ra)$ of the spin-$\f12$
representation. This corresponds in figure \ref{holonomyfig} to the
insertion of the intermediate edge labeled by the spin-$\f12$
between the legs $i$ and $j$. Dropping the index $\alpha$, this
action reads~:
\beq
\chi^{(ij)}_\alpha:\,\bigotimes_{k=1}^N
v_k
&\longrightarrow&
{\bigg{(}}
v_i\otimes v_j
\otimes
\big{(}|1,0\ra\otimes|0,1\ra-|0,1\ra\otimes|1,0\ra\big{)}
{\bigg{)}}
\bigotimes_{k\ne i,j}
v_k \nn\\
&&
={\bigg{(}}
\big{(}v_i \otimes |1,0\ra \big{)}\otimes \big{(}v_j\otimes |0,1\ra\big{)}
-\big{(}v_i \otimes |0,1\ra \big{)}\otimes \big{(}v_j\otimes |1,0\ra\big{)}
{\bigg{)}}
\bigotimes_{k\ne i,j} v_k\,.
\eeq
The crucial point is that this map is $\SU(2)$-invariant since the singlet state is invariant under $\SU(2)$. Thus it is a map between intertwiner spaces:
$$
\chi^{(ij)}_\alpha:\,
\textrm{Inv}[V_{j_i}\otimes V_{j_j} \bigotimes_k V_{j_k}]
\,\longrightarrow\,
\textrm{Inv}[\big{(}V_{j_i}\otimes V_{\f12}\big{)}\otimes \big{(}V_{j_j}\otimes V_{\f12}\big{)}
\bigotimes_k V_{j_k}].
$$
This map decomposes into four components following the decomposition into irreducible representations of the tensor products on the edges $i$ and $j$~:
$$
\big{(}V_{j_i}\otimes V_{\f12}\big{)}\otimes \big{(}V_{j_j}\otimes V_{\f12}\big{)}
\,=\,
\bigoplus_{\eta_i,\eta_j=\pm 1/2}
\big{(}V_{j_i+\eta_i}\otimes V_{j_j+\eta_j}\big{)}.
$$
This decomposition is done using the previous Clebsh-Gordan formulas
(\ref{CB1}-\ref{CB2}). The holonomy operator acts like this on both
intertwiners on the vertices $\alpha$ and $\beta$ and then glues
them by requiring that the edges $i$ and $j$ carry the same spins
$j_i+\eta_i$ and $j_j+\eta_j$ seen from both vertices $\alpha$ and
$\beta$ following the matching conditions.
Putting all the pieces together, we obtain a very simple expression
for the holonomy operator in terms of the oscillator operators
acting on both $\alpha$ and $\beta$, which can be written in terms
of $e_{ij}$ and $f_{ij}$ operators:
\be
\chi^{(ij)}
\,=\,
\f{1}{\sqrt{E_i+1}\sqrt{E_j+1}}\,
\left(
f\dag_{ij} + e_{ij} + e_{ji} +f_{ij}
\right)\,
\f{1}{\sqrt{E_i+1}\sqrt{E_j+1}}.
\label{holonomy}
\ee
The four terms correspond to the four sectors of the holonomy operators, while the inverse-squareroot factors come from the $1/\sqrt{2j+1}$ factors of the Clebsh-Gordan coefficients. Expressed as such,  the holonomy operator is clearly $\SU(2)$ invariant, Hermitian and consistent with the matching conditions. Finally, the operators $(E_k+1)$ are Hermitian and strictly positive, thus we can legitimately consider their squareroot and inverse.

This formula is surprisingly simple and expresses the holonomy
operator from loop quantum gravity in terms of the basic operators
of the $\U(N)$ formalism. It establishes a dictionnary between the
usual formulation of loop quantum gravity in terms of holonomy and
derivation (grasping) operators and our formulation in terms of the
$\u(N)$ operators $E_{ij}$ and the creation/annihilation operators
$F_{ij},F\dag_{ij}$.

\medskip

Next, we check the previous formula on bivalent intertwiners, which are the simplest states of the 2-vertex Hilbert space of spin networks. A bivalent intertwiner state is given by setting the spins to $\f J2$ on two chosen edges $i$ and $j$  and to $0$ on all other edges. This defines a unique spin network state given by the tensor product of the same singlet state of the spin-$\f J2$ representation put on both vertices $\alpha$ and $\beta$. It is also obtained by applying the holonomy operator in the spin-$\f J2$ representation on the loop $(ij)$ formed by the couple of edges. Ignoring all the other edges $k\ne i,j$, the explicit expression for this state reads:
\be
|\chi_{\f J 2}\ra
\,\equiv\,
\f{1}{J+1}
\left(
\sum_{n=0}^J(-1)^n|n,J-n\ra\otimes |J-n,n\ra
\right)^{(\alpha)}
\otimes
\left(
\sum_{p=0}^J(-1)^p|p,J-p\ra\otimes |J-p,p\ra
\right)^{(\beta)},
\ee
where the first vector in both $\alpha$ and $\beta$ terms gives the state living on the edge $i$ while the second vector defines the state on the edge $j$. This state is normalized, $\la \chi_{\f J 2}|\chi_{\f J 2}\ra=1$.
Moreover, this is a straightforward tensor product state. Thus to
derive the action of the operators $e_{ij}=\Ea_{ij}\Eb_{ij}$ and
$f_{ij}=\Fa_{ij}\Fb_{ij}$ on this state, we can drop the index
$\alpha$ (and $\beta$) and look at the action of the operators
$E_{ij}$ and $F_{ij}$ on the singlet state
$\sum_{n=0}^J(-1)^n|n,J-n\ra\otimes |J-n,n\ra$. Going back to their
expressions in terms of oscillator operators, we easily compute:
\beq
E_{ij}\,\sum_{n=0}^J(-1)^n|n,J-n\ra\otimes |J-n,n\ra &=& 0\,, \\
E_{ji}\,\sum_{n=0}^J(-1)^n|n,J-n\ra\otimes |J-n,n\ra &=& 0\,, \nn\\
F_{ij}\,\sum_{n=0}^J(-1)^n|n,J-n\ra\otimes |J-n,n\ra
&=&
-(J+1)\sum_{n=0}^{J-1}(-1)^n|n,(J-1)-n\ra\otimes |(J-1)-n,n\ra\,, \nn\\
F\dag_{ij}\,\sum_{n=0}^J(-1)^n|n,J-n\ra\otimes |J-n,n\ra
&=&
-(J+1)\sum_{n=0}^{J+1}(-1)^n|n,(J+1)-n\ra\otimes |(J+1)-n,n\ra. \nn
\eeq
Therefore, in the sum $(f\dag_{ij} + e_{ij} + e_{ji} +f_{ij})$ giving the holonomy operator, only the two terms $f_{ij}$ and $f\dag_{ij}$ changing the total area boundary contribute while the two other operator $e_{ij}$ and $e_{ji}$ vanish on the bivalent intertwiner state. This gives:
\beq
\chi^{(ij)}\,\left|\chi_{\f J 2}\right\ra
&=&
\f{1}{\sqrt{E_i+1}\sqrt{E_j+1}}\,
\left(
f_{ij} +f\dag_{ij}
\right)\,
\f{1}{\sqrt{E_i+1}\sqrt{E_j+1}}\,\left|\chi_{\f J 2}\right\ra\nn\\
&=&
\left|\chi_{\f{J-1} 2}\right\ra\,+\,\left|\chi_{\f{J+1} 2}\right\ra.
\label{holaction}
\eeq
And we recover the standard action of the holonomy operator on a single loop state. Let us point out that this action allows to easily derive the eigenvalues and eigenvectors of the holonomy operator. Indeed, the eigenvectors are given by states whose coefficients are themselves the evaluation of the characters on group elements with arbitrary but fixed  class angle $\theta$~:
\be
\chi^{(ij)}\,\sum_{J\in\N} \chi_{\f J 2}(\theta) \, \left|\chi_{\f J 2}\right\ra
\quad=\quad
\chi_{\f 1 2}(\theta)\,\sum_{J\in\N} \chi_{\f J 2}(\theta) \, \left|\chi_{\f J 2}\right\ra,
\ee
where the evaluation of the characters are:
$$
\chi_{\f J 2}(\theta)\,=\,\f{\sin(J+1)\theta}{\sin\theta},
\qquad
\chi_0(\theta)\,=\,1,
\quad
\chi_{\f 1 2}(\theta)\,=\,2\cos\theta.
$$
As we can see, the spectrum of the (one-loop) holonomy operator is bounded in absolute value by 2. Moreover, the physical interpretation of these eigenvectors is straightforward. In the limit case $\theta\arr 0$, the spin network functional corresponding to the state $\sum_J (J+1) \,|\chi_{J / 2}\ra$ is simply the distribution $\delta(g_ig_j^{-1})$ defining the flat connection state in loop quantum gravity. More generally, for an arbitrary angle $\theta$, the corresponding spin network functional is the distribution fixing the class angle of the holonomy $g_ig_j^{-1}$ to $\theta$.

\medskip

To conclude this section, we would like to give the reverse formula
of the equation \Ref{holonomy} giving the expression of the holonomy
operator in terms of the $\u(N)$ operators. To recover the
individual components $e_{ij},f_{ij},f\dag_{ij}$ from the holonomy
operator $\chi^{(ij)}$, we need to project on each of the four
sectors of the holonomy operators. This can be done by combining
$\chi^{(ij)}$ with specific insertions of $E_i$ and $E_j$. For
instance, we can select the transition $(j_i,j_j)\arr
(j_i+\f12,j_j+\f12)$ by the following combination~:
\be
\f14\bigg{(} [E_i,[E_j,\cdot]\,]+[E_i,\cdot]+[E_j,\cdot]+1 \bigg{)}
\,\chi^{(ij)} \,=\, \f{1}{\sqrt{E_i+1}\sqrt{E_j+1}}\,
f\dag_{ij}
\,
\f{1}{\sqrt{E_i+1}\sqrt{E_j+1}}\,.
\ee
We can adapt this formula to select the other transitions and select the other components of the holonomy operator:
\beq
\f14\bigg{(} [E_i,[E_j,\cdot]\,]-[E_i,\cdot]-[E_j,\cdot]+1 \bigg{)}
\,\chi^{(ij)} &=& \f{1}{\sqrt{E_i+1}\sqrt{E_j+1}}\, f_{ij} \,
\f{1}{\sqrt{E_i+1}\sqrt{E_j+1}}\,,\\
\f14\bigg{(} -[E_i,[E_j,\cdot]\,]+[E_i,\cdot]-[E_j,\cdot]+1 \bigg{)}
\,\chi^{(ij)} &=& \f{1}{\sqrt{E_i+1}\sqrt{E_j+1}}\, e_{ij} \,
\f{1}{\sqrt{E_i+1}\sqrt{E_j+1}}\,.\nn
\eeq
This concludes the dictionary between the holonomy operators and the  operators of the $\U(N)$ formalism.

\medskip

Now that we have explored in detail the algebraic structure of the
space of spin network states on the 2-vertex graph, we will look at
the $\U(N)$-invariant operators acting in our Hilbert space $\dcH$
and will define $\U(N)$-invariant dynamics that evolve homogeneous
states. This will allow us to make a link between our formalism and the
standard framework of loop quantum cosmology.

\section{Dynamics on the 2-Vertex Graph}

\subsection{The algebra of $\U(N)$-invariant operators}

In the previous section, we have introduced the $\U(N)$-invariant space $\cHi$ of homogeneous states spanned by basis vectors $|J\ra$ labeled by solely the total boundary area. Now, we would like to study the structure of  $\U(N)$-invariant operators that will act on this space $\cHi$ of invariant states.

The most obvious $\U(N)$-invariant operator is the total boundary area operator $E$ itself. It is defined as $E=\Ea=\Eb$ on the space $\dcH$ of spin network states satisfying the matching condition. It is direct to check that it commutes with the $\u(N)$ generators:
\be
[\cE_{ij},\Ea]=[\Ea_{ij},\Ea]
\,=\,
0
\,=\,
-[\Eb_{ji},\Eb]
=[\cE_{ij},\Eb].
\ee
This total area operator is clearly diagonal in the basis $|J\ra$,
$$
E\,|J\ra
\,=\,
2J\,|J\ra.
$$
Now, we need operators that could create dynamics on the space $\cHi$ by inducing transitions between states with different areas. To this purpose, we use the operators $e_{ij},f_{ij}$ and introduce the unique linear combinations that are $\U(N)$-invariant:
\be
e\equiv\sum_{ij}e_{ij}=\sum_{ij}\Ea_{ij}\Eb_{ij},\qquad
f\equiv\sum_{ij}f_{ij}=\sum_{ij}\Fa_{ij}\Fb_{ij}.
\ee
They obviously commute with the matching conditions since each
operator $e_{ij}$ and $f_{ij}$ does. Moreover, it is straightforward to show
that they are also invariant under $\U(N)$ transformations:
\be
[\cE_{ij},e]=0\,,\qquad [\cE_{ij},f]=0\,,\qquad
[\cE_{ij},f\dag]=0\,.
\ee
These three operators $e,f,f\dag$ seem to be the only
$\U(N)$-invariant operators which are quadratic in $E$ and/or $F$
and satisfy the matching conditions (the action of the operators on
both vertices has to be the same in order to preserve these
conditions, so it is necessary for such quadratic operators not to
mix the operators $E_{ij}$, $F_{ij}$ and $F\dag_{ij}$ in the same
term).

Using the quadratic constraints (\ref{constraint1}-\ref{constraint3})
satisfied by the operators $E_{ij}$ and $F_{ij}$, we can show that $e$, $f$ and $f\dag$ form a 
simple algebra:
\beq
\left[e,f\right]&=&-2f(E+N-3)=-2(E+N-1)f\,,\label{comm1}\\
\left[e,f^{\dagger}\right]&=&2f^{\dagger}(E+N-1),\,\label{comm2}\\
\left[f,f^{\dagger}\right]&=&4(E+N)(e+2(E+N-1))\,.
\label{comm3}
\eeq
%
%
Looking at these commutators, it seems natural to introduce a shifted operator $\tl{e}\equiv e+2(E+N-1)$. Then the algebra reads:
\beq
\left[\te,f\right]&=&-2(E+N+1)f\,,\\
\left[\te,f^{\dagger}\right]&=&2f^{\dagger}(E+N+1),\,\nn\\
\left[f,f^{\dagger}\right]&=&4(E+N)\te\,.\nn
\eeq
Written as such, it resembles to a $\sl_2$ Lie algebra up to the
factors in $E$, which is an operator and not a constant. This is
very similar to the $\sl_2$ algebra $\Ea_\rho,\Fa_\bz,\Fa_\bz{}\dag$
defined in \cite{un3} and used to build the $\U(N)$ coherent
states\footnotemark.
\footnotetext{ In the previous work \cite{un3}, operators
$\Ea_\rho,\Fa_\bz,\Fa_\bz{}\dag$ were constructed by contracting the
operators $\Ea_{ij},\Fa_{ij},\Fa_{ij}{}\dag$ with the Hermitian
matrix $\rho_{ij}$ and the antisymmetric  matrix $\bz_{ij}$ both
defined in terms of the set of $N$ given spinors
$\{(z^0_i,z^1_i)\}_{i=1..N}$~:
\beq
\Ea_\rho\equiv \sum_{ij}\rho_{ij}\Ea_{ij},&\qquad&
\rho_{ij}\equiv {z^0_i}\overline{z^0_j}+{z^1_i}\overline{z^1_j}\,, \nn\\
\Fa{}\dag_\bz\equiv \sum_{ij}\bz_{ij}\Fa_{ij}{}\dag,&\qquad&
\bz_{ij}\equiv z^0_iz^1_j-z^0_jz^1_i \,.\nn
\eeq
Then the three  operators $\Ea_{ij},\Fa_{ij},\Fa_{ij}{}\dag$ form a
$\sl_2$ algebra. In our framework, working on the two-vertex graph,
we have replaced these two classical matrices $\rho$ and $\bz$ by
the operators acting on the $\beta$-vertex, $\rho_{ij}\arr\Eb_{ij}$
and $\bz_{ij}\arr\Fb_{ij}{}\dag$. This basically amounts to replace
the spinor labels by the annihilation operators of the
$\beta$-vertex,
$(z^0_i,z^1_i)\arr(a^{(\beta)\dagger}_i,b^{(\beta)\dagger}_i)$. From
this point of view, our 2-vertex framework generalizes the algebraic
structure of the single intertwiner case and it is natural to find
similar algebras. }
%
%
Since $f,f\dag$ play the role of lowering and raising  operators in this algebra, we can use $f\dag$ as a creation operator. Thus we introduce the states:
\be
|J\ra_{un}\equiv f^{\dagger J}|0\ra
\,=\,
\left(\sum_{ij}\Fa_{ij}{}\dag\Fb_{ij}{}\dag\right)^J\,|0\ra,
\ee
where the index ${un}$ stands for un-normalized (or for $\U(N)$ following the reader's preference). Since both the operator $f\dag$ and the vacuum state $|0\ra$ are $\U(N)$-invariant, it is clear that the states $|J\ra_{un}$ are also invariant under the $\U(N)$-action. Moreover, it is easy to check that they are eigenvectors of the total area operator:
$$
E\,|J\ra_{un}=2J\,|J\ra_{un},
$$
so that they provide a basis for our Hilbert space $\cHi$ of homogeneous states.
Then, using the commutators (\ref{comm1}-\ref{comm3}), we can
compute the action of the operators $e$, $f$ and $f\dag$ on these
states\footnotemark:
\beq
e|J\ra_{un} &=& 2J(N+J-2)|J\ra_{un}\,,\\
\te|J\ra_{un} &=& 2(J+1)(N+J-1)|J\ra_{un}\,,\nn\\
f|J\ra_{un} &=& 4J(J+1)(N+J-1)(N+J-2)|J-1\ra_{un}\,,\nn\\
f\dag|J\ra_{un} &=&|J+1\ra_{un}\,.\nn
\eeq
As expected, the states $|J\ra_{un}$ diagonalize $e$, while $f\dag$ and $f$ act respectively as creation and annihilation operators.
\footnotetext{ We use the commutation relations  of $e$ and $f$ with
$f\dag$ to compute their action:
$$
e|J\ra_{un}=ef\dag{}^J|0\ra_{un} \,=\,\sum_{k=0}^{J-1}
f\dag{}^{J-1-k}[e,f\dag]f\dag{}^k|0\ra \,=\,2\sum_{k=0}^{J-1}
(2k+N-1)\, f\dag{}^J|0\ra \,=\,2J(N+J-2)\, f\dag{}^J|0\ra,
$$
$$
f|J\ra_{un}=ff\dag{}^J|0\ra \,=\,\sum_{k=0}^{J-1}
f\dag{}^{J-1-k}[f,f\dag]f\dag{}^k|0\ra \,=\,4\sum_{k=0}^{J-1}
(2k+N)(2k(k+N-2)+2(2k+N-1)))\, f\dag{}^{J-1}|0\ra.
$$}
The action of $e$ provides a consistency check. Indeed, taking into account the $\u(N)$ constraints $\Ea_{ij}-\Eb_{ji}=0$ imposed on our invariant space $\cHi$, the operator $e$ is simply equal to the quadratic $\U(N)$-Casimir on that Hilbert space~:
\be
e|J\ra_{un}
\,=\,
\sum_{ij}\Ea_{ij}\Eb_{ij}|J\ra_{un}
\,=\,
\sum_{ij}\Ea_{ij}\Ea_{ji}|J\ra_{un}
\,=\,
\Ea\left(\f \Ea 2+N-2\right)|J\ra_{un}
\,=\,
2J(J+N-2)|J\ra_{un},
\ee
where the value of the Casimir operator was given earlier in \Ref{casimir}. Similarly, we can express the shifted operator $\te$ as a polynomial in the total area operator $E$ on the invariant space $\cHi$~:
\be
\te\,|\psi\ra
\,=\,
\f12(E+2)\,(E+2(N-1))\,|\psi\ra,\quad
\forall\psi\in\cHi.
\ee
Using this action, we compute the norm of the states $|J\ra_{un}$ by recursion:
\be
{}_{un}\la J|J\ra_{un}
\,=\,
\la 0|f^{J-1}\,f|J\ra
\,=\,
4J(J+1)(N+J-1)(N+J-2)\,{}_{un}\la {J-1}|J-1\ra_{un},
\ee
which leads us to the scalar product:
\be
{}_{un}\la J|J\ra_{un}
\,=\,
2^{2J}J!(J+1)!\f{(N+J-1)!(N+J-2)!}{(N-1)!(N-2)!}
\,=\, 2^{2J}\,(J!(J+1)!)^2\,D_{N,J},
\ee
in terms of the dimension of the intertwiner space $\cH_N^{(J)}$
given in \Ref{dimNJ}. Thus we can define normalized basis states:
\be
|J\ra\,\equiv\,
\f{1}{2^J\,J!(J+1)!\,\sqrt{D_{N,J}}}\,|J\ra_{un},
\ee
on which the action of all $e,f,f\dag$ operators is always quadratic in $J$:
\beq
\te\,|J\ra &=& 2(J+1)(N+J-1)|J\ra\,,\\
f\,|J\ra &=& 2\sqrt{J(J+1)(N+J-1)(N+J-2)}\,|J-1\ra,\nn\\
f\dag\,|J\ra &=&2\sqrt{(J+1)(J+2)(N+J)(N+J-1)}\,|J+1\ra.\nn
\eeq

\medskip

From here, we see that it is possible to introduce renormalized operators that truly form a $\sl_2$ algebra. We define the new operators:
\beq
Z&\equiv&
\f{1}{\sqrt{E+2(N-1)}}\,\te\,\f{1}{\sqrt{E+2(N-1)}}\,,\nn\\
X_-&\equiv&
\f{1}{\sqrt{E+2(N-1)}}\,f\,\f{1}{\sqrt{E+2(N-1)}}\,,\\
X_+&\equiv&
\f{1}{\sqrt{E+2(N-1)}}\,f\dag\,\f{1}{\sqrt{E+2(N-1)}}\,.\nn
\eeq
First, the inverse square-root is well-defined since $E+2(N-1)$ is Hermitian and strictly positive as soon as $N\ge 2$. Moreover, these operators are still $\U(N)$-invariant since $E$ is invariant too. We also  have the Hermiticity relations, $Z\dag=Z$ and $X_-\dag=X_+$. It is fairly easy to compute the action of these renormalized operators on our $|J\ra$ basis states:
\beq
\label{sl2action}
Z\,|J\ra &=& (J+1)\,|J\ra\,,\\
X_-\,|J\ra &=& \sqrt{J(J+1)}\,|J-1\ra,\nn\\
X_+\,|J\ra &=&\sqrt{(J+1)(J+2)}\,|J+1\ra.\nn
\eeq
These operators clearly form a $\sl(2,\R)$ Lie algebra:
\be
[Z,X_\pm]=\pm X_\pm,\qquad
[X_+,X_-]=-2Z.
\ee
Furthermore, the action \Ref{sl2action} corresponds to the irreducible unitary representation of $\sl(2,\R)$ with vanishing Casimir $\cQ=-Z^2+(X_-X_++X_+X_-)/2=0$. This irrep belongs to the positive discrete series of irreps of $\sl(2,\R)$, which have a spectrum bounded from below for the operator $Z$ (the interested reader can find more details on the unitary representations of $\sl(2,\R)$ in \cite{su11}).

This shows two important points:
\begin{itemize}

\item The algebraic structure of the $\U(N)$-invariant space $\cHi$ of homogeneous states is clear: it forms an irreducible unitary representation of $\sl(2,\R)$. The basis vectors $|J\ra$ can be obtained by iterating the action of the creation/raising operator $f\dag$  (or $X_+$) on the vacuum state $|0\ra$.

\item This algebraic structure does not depend at all on the number of edges $N$. Therefore, while working on homogeneous states, $N$ might have a physical meaning but it is not a relevant parameter mathematically. On the other hand, we expect it to become highly relevant when leaving the $\U(N)$-invariant subspace and studying inhomogeneities.

\end{itemize}
Since the algebra of the operators $Z,X_\pm$ is much simpler than
the algebra of the operators $\te,f,f\dag$, it is also convenient to
give the reverse formula and write $\te,f,f\dag$ in terms of the
$\sl_2$ operators. Using that $Z$ is simply related to the total
area $E$ on the invariant space $\cHi$,
$$
Z=\f E 2 +1,\qquad
E +2(N-1)
= 2(Z+N-2),
$$
we have the inverse relations, which hold on the Hilbert space $\cHi$:
\beq
\te &=& 2Z(Z+N-2)\,,\\
f &=& 2\sqrt{Z+N-2}\, X_- \,\sqrt{Z+N-2}\,, \nn\\
f\dag &=& 2\sqrt{Z+N-2}\, X_+ \,\sqrt{Z+N-2}\,.
\eeq

\medskip

Finally, to conclude this section, we introduce some further renormalized operators:
\be
\f 1 {\sqrt{\te}}\,\te\,\f 1 {\sqrt{\te}}=\id,\qquad
\f 1 {\sqrt{\te}}\,f\,\f 1 {\sqrt{\te}},\qquad
\f 1 {\sqrt{\te}}\,f\dag\,\f 1 {\sqrt{\te}}.
\ee
These operators obviously still commute with the $\U(N)$ action. They are also well-defined since $\te$ is Hermitian and strictly positive. Their action on the basis states is the simplest, since they generate straightforward translations between states of different areas:
\beq
\f 1 {\sqrt{\te}}\,f\,\f 1 {\sqrt{\te}} \,|J\ra
&=& |J-1\ra\,,\quad\forall J\ge 1, \\
\f 1 {\sqrt{\te}}\,f\dag\,\f 1 {\sqrt{\te}} \,|J\ra
&=& |J+1\ra\,.\nn
\eeq
Nevertheless, we will not use a lot these renormalized operators and we will focus below on dynamical operators built from the original operators $\te,f,f\dag$.

\subsection{``Black Hole" Creation Operator}
\label{UNstate}

Before studying the dynamics of the spin network states on our 2-vertex graph, we would like to explain further the physical interpretation of our homogeneous states $|J\ra\,\propto\,(f\dag)^J\,|0\ra$. In fact, these states are maximally entangled states between $\alpha$ and $\beta$. More precisely, let us consider the density matrix corresponding to this pure state and define the reduced density matrix on the system $\alpha$ obtained by tracing out the system $\beta$~:
\beq
&&
\rho
\,\equiv\,
|J\ra\la J|
\propto
(f\dag)^J\,|0\ra\la 0| f^J,\quad\tr\rho=1,\\
&&
\rho_\alpha
\,\equiv\,
\tr_\beta\,\rho
\,\equiv\,
\tr_{\cH_N^{(J)(\beta)}}\,\rho\,.
\nn
\eeq
The main point is that the reduced density matrix on $\alpha$ is the totally mixed state on the Hilbert space of $N$-valent intertwiners with fixed total area $J$,
$$
\rho_\alpha\propto\id_\alpha=\id_{\cH_N^{(J)}}\,.
$$
To prove this, we first check that the density matrix $\rho_\alpha$ commutes with the $\U(N)$-action on $\cH_N^{(J)(\alpha)}$. This comes directly from the fact that the full density matrix is by definition invariant under the $\U(N)$-action acting on both vertices,
$\rho=(U\otimes\bar{U})\,\rho\,(U^{-1}\otimes\bar{U}^{-1})$ as matrices on $\cH_N^{(J)(\alpha)}\otimes\cH_N^{(J)(\beta)}$. Taking this into account, we have~:
\be
\forall U\in\U(N),\quad
U\rho_\alpha U^{-1}
\,=\,
\tr_\beta\,(U\otimes \id)\rho(U^{-1}\otimes \id)
\,=\,
\tr_\beta\,(\id\otimes \bar{U}^{-1})\rho(\id\otimes \bar{U})
\,=\,
\tr_\beta\,\rho
\,=\,
\rho_\alpha\,,
\ee
since the trace on $\cH_N^{(J)(\beta)}$ is obviously invariant under unitary since it carries an (irreducible) representation of $\U(N)$. Once we have checked that $\rho_\alpha$ commutes with unitary transformations, we can conclude by Sch\" ur lemma that it is necessarily proportional to the identity on the intertwiner space $\cH_N^{(J)(\alpha)}$ since this space carries (once again) an irreducible representation of $\U(N)$. Thus, taking care of the normalization, we conclude that:
\be
\rho_\alpha
\,=\,
\f1{D_{N,J}}\,\id_{\cH_N^{(J)}}\,.
\ee
This means that the we have a mixed state with maximal entropy on the system $\alpha$ and that our initial pure state is maximally entangled.

In previous work \cite{un2}, it was shown that the entropy of this state, given by the logarithm of the dimension of the intertwiner space $\cH_N^{(J)}$, grows linearly with the total boundary area $J$ in the large scale limit, and thus behaves holographically. It is tempting to call this a ``black hole" state in our very simplistic framework, since we have a pure state on the coupled system with a ``maximal ignorance" state for each of the two sub-systems. From this perspective, we can dub our operators $(f\dag)^J$ as ``black hole" creation operators.

To insist on this point of view, we remind that it has been recently argued from various perspectives that the  Hilbert space of states for a (quantum) black hole in loop quantum gravity is the space of intertwiners with the same fixed total area \cite{bh1,bh1bis,bh2,bh3,bh4}. This supports our claim here that the induced reduced density matrix $\rho_\alpha$ by the pure homogeneous state $|J\ra$ represents a physical configuration similar to a black hole.

%

\subsection{An Ansatz for Dynamics}

Now, we would like to study the dynamics on this 2-vertex graph and propose the simplest
$\U(N)$-invariant ansatz for a Hamiltonian operator:
\be
H
\,\equiv\,
\lambda\te+ (\sigma f +\bar{\sigma} f\dag).
\ee
As explained above, the operator $\te$ does not affect the total
boundary area, $[E,\te]=0$, while the operators $f$ and $f\dag$
respectively shrink and increase this area, $[E,f]=-2f$ and
$[E,f\dag]=+2f\dag$. The coupling $\lambda$ is real while $\sigma$
can be complex a priori, so that the operator $H$ is Hermitian.
We can relate this Hamiltonian operation to the action of holonomy
operators acting on all the loops of the 2-vertex graph using the
formula \Ref{holonomy}. From this point of view, our proposal is
very similar to the standard ansatz for the quantum Hamiltonian
constraint in loop quantum gravity \cite{thomas1} and loop quantum
cosmology (e.g. \cite{KLP,lqc_spectrum3}).
%

This Hamiltonian is quadratic in $J$
since it is of order 4 in the oscillators, and we can give its explicit action on the basis states of $\cHi$:
\beq
H|J\ra
&=&
\sigma \,2\sqrt{J(J+1)(N+J-1)(N+J-2)}\,|J-1\ra\nn\\
&&+\lambda\,2(J+1)(N+J-1)\,|J\ra\nn\\
&&+ \bar{\sigma}\,2\sqrt{(J+1)(J+2)(N+J)(N+J-1)}\,|J+1\ra\,.
\eeq
We will study the properties, especially its spectral properties, in
detail in the next section. This operator is particularly
interesting because it allows a direct link with the framework of
loop quantum cosmology, as we will discuss in section \ref{toLQC}.
In the following, we will restrict ourselves to the case of real
couplings, $\sigma\in\R$.

Looking for the eigenvalues and eigenvectors of this Hamiltonian,
$H\,\sum_J \alpha_J |J\ra\,=\,2\beta\,\sum_J \alpha_J |J\ra$, is
translated into a second order recursion relation\footnotemark \,for
the coefficients $\alpha_J$~:
\beq
\forall J\ge 1,\quad&
\sigma\sqrt{(J+1)(J+2)(N+J)(N+J-1)}\,\alpha_{J+1}
&+ \lambda\left[(J+1)(N+J-1)-\beta\right]\,\alpha_J\nn\\
&&
+\sigma\sqrt{J(J+1)(N+J-1)(N+J-2)}\,\alpha_{J-1}=0.
\eeq
\footnotetext{
This second order recursion relation is very similar  to the recursion relation for the $\{6j\}$-symbol of the recoupling theory for $\SU(2)$ \cite{SG,maite}. Such a relation can almost be expected since both our states $|J\ra$ and the $\{6j\}$-symbol define distributions on the space of $\SU(2)$ intertwiners. Furthermore, directly comparing by hand the  two recursion relations, it is possible to have an almost perfect match between the coefficients of the recursion relations by setting $N=1$ and considering isosceles tetrahedra for the $\{6j\}$-symbol. From this relation, we expect three regimes for the coefficients $\alpha_J$ corresponding to the three asymptotic regimes of the $\{6j\}$-symbol: an oscillatory phase (corresponding to true Euclidean tetrahedra), an exponential phase (which can be mapped to tetrahedra embedded in the 2+1d Minkowskian space) and a critical regime (which corresponds mathematically to tetrahedra with vanishing squared-volume).
}
We will study in detail the solutions to this equation and the
resulting spectral properties of the operator $H$ in the next
section \ref{spectral}. Nevertheless, we can already make some
simple comments about it. First, although it is a second order
recursion relation, the initial equation at $J=0$ determines the
coefficient $\alpha_1$ in terms of $\alpha_0$. Thus we only need one
initial condition $\alpha_0$ and the space of solutions to this
recursion relation is one-dimensional once $\beta$ is fixed.  The
second remark is that since it is a second order recursion relation,
we expect the coefficients $\alpha_J$ to be oscillatory, at least
for large $J$'s. Considering the leading order in $J$ in the
asymtptotics, it is fairly easy to see that the coefficients of the
recursion relation only depends on the couplings $\lambda$ and
$\sigma$ so the frequency of the oscillations does not depend on $N$
and the eigenvalue $\beta$ but is expressed only in terms of
$\lambda$ and $\sigma$~:
\be
\cos\om =-\lambda/2\sigma.
\ee
This means that we have a priori three dynamical regimes, depending
on the value of the couplings. Taking $\lambda > 0$ for the sake of
simplicity, we distinguish the three following cases:
\begin{enumerate}

\item the {\it oscillatory regime} when $|\sigma|>\lambda /2$~: the frequency $\om$ is real and we expect the coefficients $\alpha_J$ to oscillate as $\exp(i\om J)$  as $J$ grows large.

\item the {\it discrete regime} when $|\sigma|<\lambda /2$~: the frequency $\om$ is purely imaginary and the coefficients $\alpha_J$ to diverge as $J$ grows to infinity. Actually, they always diverge except for a discrete set of choices of eigenvalues $\beta$. Then the spectrum of $H$ turns out to be discrete and positive.

\item the {\it critical regime} when $\sigma =\pm\lambda /2$~:
the frequency vanishes when $\sigma =-\lambda /2$ and is equal to $\pi$ when $\sigma =+\lambda /2$. In the case $\sigma =-\lambda /2$, our model is very similar to the flat FRW model of loop quantum cosmology (e.g. \cite{merce}) and we expect similar asymptotics for our eigenstates. In particular, we will show that the spectrum of $H$ is continuous, real, and positive.

\end{enumerate}
We will study the details of these three regimes in the next section.
Moreover, we will discuss later, in section \ref{constant}, the relation between our couplings $\lambda$ and $\sigma$ and the cosmological constant, comparing these three regimes respectively to the three cases $\Lambda<0$, $\Lambda>0$ and finally $\Lambda=0$.

\medskip

The ansatz given above is the most general U(N) invariant
Hamiltonian (allowing only elementary changes in the total area), up
to a renormalization by a $E$-dependent factor. Therefore, we can
also propose renormalized Hamiltonian operators based on the
renormalized operators considered in the previous section. For
instance, we define both:
\be
\hh\,\equiv\,
\f{1}{\sqrt{E+2(N-1)}}\,H\,\f{1}{\sqrt{E+2(N-1)}}
\,=\,
\lambda Z+(\sigma X_-+\bar{\sigma} X_+)\quad\in\,\sl_2\,,
\ee
\be
h\,\equiv\,
\f1{\sqrt{\te}}\, H\,\f1{\sqrt{\te}}
= \lambda\id +\f1{\sqrt{\te}}\,(\sigma f +\bar{\sigma} f\dag)\,\f1{\sqrt{\te}}.
\ee
Let us underline the fact that we study the action of these Hamiltonian operators $H$, $\hh$ and $h$ on the $\U(N)$-invariant space $\cHi$, nevertheless they are generally well-defined on the whole space of spin network states $\dcH$.
Let us also remind the fact that the parameter $N$ giving the number of edges of the graph has completely disappeared from the action of these renormalized Hamiltonians $\hh$ and $h$: the dynamics on the homogeneous sector does not depend mathematically on $N$.

In the context of loop quantum gravity or loop quantum cosmology, these various (re)normalization for the Hamiltonian operator would correspond to different densitized versions of the Hamiltonian operator. Comparing to loop quantum cosmology e.g. \cite{KLP}, our initial Hamiltonian operator seems to correspond to the {\it  evolution operator} $\hat{\Theta}$ while our renormalized Hamiltonian $\hh$ corresponds to the gravitational part of the LQC Hamiltonian constraint $\hat{C}_{\textrm{grav}}$.

The main characteristic of $\hh$ is that its coefficients are linear
in the variable $J$ at leading order and, more importantly, it is an
element in the Lie algebra $\sl_2$. Thus we know its spectral
properties from the representation theory of $\sl_2$. More
precisely, for a real couplings $\lambda,\sigma\in\R$, the generator
$\hh=\lambda Z+(\sigma X_-+\bar{\sigma} X_+)$ lies in the Lie
algebra $\sl(2,\R)\sim\su(1,1)$. The (Lorentzian) norm of this
3-vector,
\be
||\hh||^2
\,=\,
\lambda^2-4|\sigma|^2,
\ee
determines its type. There are three cases~:
\begin{enumerate}

\item If $\lambda^2>4|\sigma|^2$, $\hh$ is time-like, i.e it generates a space rotation in $\SU(1,1)$ and it is conjugated to $Z$ by a $\SU(1,1)$ transformation (up to a numerical factor). In this case, the spectrum of $\hh$ is discrete. More precisely, since the spectrum of $Z$ is $\N^*$, then the spectrum of $\hh$ is ${\rm sign}(\lambda)\,||h||\,\N^*$.

\item If $\lambda^2<4|\sigma|^2$, $\hh$ is space-like, i.e it generates a boost in $\SU(1,1)$, its spectrum is continuous and covers the whole real line.

\item If $\lambda^2=4|\sigma|^2$, $\hh$ is null-like and its spectrum is also continuous and of the same sign than $\lambda$.

\end{enumerate}

For more details on the spectral analysis of $\su(1,1)$ algebra elements and their eigenvectors, the interested reader can refer to \cite{su11davids}. The important point to underline here is that the three coupling regimes for the renormalized Hamiltonian $\hh$ are exactly the same as for the original Hamiltonian $H$.
This is very similar to the interplay between the two operators in LQC, the {\it  evolution operator} $\hat{\Theta}$ and the gravitational contribution to the Hamiltonian constraint $\hat{C}_{\textrm{grav}}$ (see e.g. \cite{KLP}).

The advantage of this $\sl_2$ Hamiltonian operator $\hh$ over the initial Hamiltonian $H$ is that its properties are much better known. For instance, its spectral properties are well-established and we know its exact discrete spectrum in the weak coupling case $\lambda^2>4|\sigma|^2$. Also the exponentiation of Lie algebra elements is under control and we know that $\hh$ will generate finite $\SU(1,1)$ transformations. Nevertheless, we will also analyze the properties of $H$ since this choice of Hamiltonian seems to correspond to the operator used in LQC.

\medskip

The main feature of the last operator $h$ is that its coefficients are of order 0 in the area $J$. Actually, its action is truly simple:
\be
h|J\ra\,=\,
\lambda|J\ra \,+ \sigma|J-1\ra+\bar{\sigma}|J+1\ra,\qquad
h|0\ra\,=\,
|0\ra \,+\bar{\sigma}|1\ra.
\ee
This operator has properties very similar to the holonomy operator
$\chi^{(ij)}$ whose action was given in \Ref{holaction}. Restricting
ourselves to the case of a real coupling $\sigma\in\R$ for the sake
of simplicity, the Hamiltonian $h$ has a bounded spectrum and its
eigenvectors (in the generalized sense) are given by plane waves
(for more details, see the appendix)~:
\be
h\,|\om\ra
\,=\,
(\lambda+2\sigma\cos\om) \,|\om\ra,
\qquad\textrm{with}\quad
|\om\ra
\,\equiv\,
\sum_{J\in\N} \sin\big{[}(J+1)\om\big{]}\,|J\ra.
\ee
The eigenvalue $\beta=\lambda+2\sigma\cos\om$ corresponding to the frequency $\om$ is obviously bounded by the extremal values $\lambda\pm 2\sigma$. This operator is less interesting than $H$ and $\hh$, and is not relevant to the relationship between our framework to loop quantum cosmology.

\subsection{Solving the 2-Vertex Model}
\label{spectral}

In this section, we study in more details the spectral properties of
the Hamiltonian $H$ depending on the couplings $\lambda$ and
$\sigma$. In particular, we will establish the existence of the
three regimes discussed above. To this purpose, we will use the
techniques used in LQC to study the properties of the gravitational
part of the Hamiltonian constraint
\cite{lqc_spectrum1,lqc_spectrum2,lqc_spectrum3}. In the following,
we will assume that $\lambda>0$ for the sake of simplicity.

We start by checking the positivity of our operator (which we rescale by a factor 2):
\be
H|J\ra \,=\,
\lambda \vphi(J)\,|J\ra
+\sigma \psi(J)\,|J-1\ra
+\bar{\sigma} \psi(J+1)\,|J+1\ra\,,
\quad\textrm{with}\qquad
\left|
\begin{array}{lcl}
\vphi(J)&=&
(J+1)(N+J-1)\\
\psi(J) &=&
\sqrt{J(J+1)(N+J-1)(N+J-2)}
\end{array}
\right.
\label{hamiltonian}
\ee
We consider the diagonal part of this operator as our free Hamiltonian $H_0\,|J\ra \,\equiv\, \vphi(J)\,|J\ra$. Actually it is  simply our operator $\te$ up to a factor 2. Then the off-diagonal components triggering the transitions to the $|J\pm1\ra$ states are considered as the interaction:
\be
H=\lambda H_0+I\quad
\textrm{with}\qquad
I\,|J\ra\,=\,\sigma \psi(J)\,|J-1\ra
+\bar{\sigma} \psi(J+1)\,|J+1\ra\,.
\ee

\begin{res}
For  a weak coupling, $|\sigma|\le \lambda/2$, the Hamiltonian H is
positive, $\forall v\ne 0\, \la v|H|v\ra> 0$\,.
\end{res}

\begin{proof}
Writing the vector $v$ in the fundamental basis, $v=\sum_J v_J \,|J\ra$, we compute explicitly the action of the Hamiltonian:
$$
\la v|H|v\ra\,=\,
\sum_J \lambda \vphi(J) |v_J|^2 + \sigma \psi(J)\overline{v_{J-1}}v_J
+ \bar{\sigma} \psi(J+1)\overline{v_{J+1}}v_J
\,.
$$
Then to show the positivity, since $\vphi(J)>0$ and thus the free term is always positive, we just need to show that the interaction term is smaller in than this free term.
First we use  the obvious inequality, $|\overline{v_{J+1}}v_J|\le (|v_J|^2+|v_{J+1}|^2)/2$, which means that:
$$
\left|
\sum_J \sigma \psi(J)\overline{v_{J-1}}v_J
+ \bar{\sigma} \psi(J+1)\overline{v_{J+1}}v_J
\right|
\,\le\,
|\sigma|\,\sum_J (\psi(J)+\psi(J+1))\,|v_J|^2.
$$
Now, we only have to check that $0\le(\psi(J)+\psi(J+1))< 2\vphi(J)$ holds for all $J$'s. This is straightforward to show by squaring this inequality. Therefore, we can conclude that, as soon as $2|\sigma|\le \lambda$, the Hamiltonian is positive. On the other hand, it is easy to see that as soon as $2|\sigma|> \lambda$, then the expectation value $\la v|H|v\ra$ can take arbitrary real values.

\end{proof}

\begin{res}
We can bound the interaction term by the free Hamiltonian:
\be
\forall v,\quad
||I\, v||\le {2|\sigma|}\,||H_0\,v||\,+\, {|\sigma|}\,||v||\,.
\ee
\end{res}

\begin{proof}
We follow the same steps as in the previous proof:
\beq
||I\, v||^2
&=&
\bigg{|}\bigg{|}\sum_J \sigma \psi(J)v_J\,|J-1\ra+\sum_J \bar{\sigma} \psi(J+1)v_J\,|J+1\ra\bigg{|}\bigg{|}^2 \nn\\
&\le &
2|\sigma|^2\sum_J (|\psi(J)|^2+|\psi(J+1)|^2)\,|v_J|^2,
\eeq
where we used that for an arbitrary couple of vectors $v,w$, we have
the norm inequality $||v+w||^2\le 2(||v||^2+||w||^2)$. Then we
compute explicitly:
$$
\psi(J)^2+\psi(J+1)^2
\,=\,
2(J+1)(N+J-1)(J^2+NJ+N)
\,=\, 2\vphi(J)(\vphi(J)+1)\,.
$$
Thus, inserting this formula back in the previous inequality, we get:
\beq
||I\, v||^2
&\le&
4|\sigma|^2\left(||H_0\,v||^2+\la v|H_0\,v\ra\right)\nn\\
&\le&
4|\sigma|^2\left(||H_0\,v||^2+ ||v||\,||H_0\,v||\right)\nn\\
&\le&
4|\sigma|^2\left(||H_0\,v||+ \f12||v||\right)^2\,,
\eeq
which allows us to obtain the bound.

\end{proof}

Since the operator $H_0$ is obviously essentially self-adjoint, and since $H=\lambda H_0+I$, we can apply the previous inequality in the small coupling case,
$$
|\sigma|\le \f\lambda 2
\quad\Rightarrow \quad
||I\,v||
\,\le\,
||\lambda H_0\,v||\,+\, {|\sigma|}\,||v||\,,
$$
and conclude that:

\begin{res}
The Hamiltonian $H$ is essentially self-adjoint as soon as $2|\sigma|\le\lambda$.
\end{res}

This allows to prove the existence of the {\it discrete regime} when $2|\sigma|<\lambda$~:

\begin{res}
When $2|\sigma|<\lambda$, the spectrum of $H$ is discrete and strictly positive.
\end{res}

\begin{proof}

Assuming that $2|\sigma|<\lambda$, we consider the following
decomposition of the Hamiltonian:
$$
H=(\lambda-2|\sigma|)H_0
\,+\,
\big(2|\sigma|H_0+I\big).
$$
Using the result 1, we know that the operator $(2|\sigma|H_0+I)$ is positive. Thus we have the operator inequality $0\le (\lambda-2|\sigma|)H_0 \le H$.

Finally, we use the facts that $H$ is essentially self-adjoint for our choice of coupling and that $(\lambda-2|\sigma|)H_0$ has a discrete positive spectrum and we apply the lemma in \cite{lqc_spectrum1} (p.16) to conclude that $H$ has a discrete spectrum.

We do not repeat the rigorous proof of this lemma here. Nevertheless, the intuitive idea is that the subspace generated by eigenvectors of $H$ with eigenvalue inferior to an arbitrary value $\beta_m$ is automatically included in the subspace of of eigenvectors of $(\lambda-2|\sigma|)H_0$ with eigenvalue inferior to the same max value $\beta_m$. Since the latter has a finite dimension, the first also does.

\end{proof}

Now we turn to the {\it critical regime} when $2|\sigma|=\lambda$.
For the sake of simplicity, we fix $\lambda=2$, then $\sigma$ is on
the unit circle, $\sigma=\exp(-i\theta)$ with an arbitrary angle
$\theta\in[0,2\pi]$. We will write $H_c$ for the Hamiltonian
operator in this critical regime. We already know from the previous
analysis that $H_c$ is essentially self-adjoint and strictly
positive. We will not provide here rigorous proofs about its
spectral properties. We will present instead approximative
asymptotics for its eigenvectors. For a more rigorous approach, one
could use techniques similar than the ones used in loop quantum
cosmology, performing for instance a Fourier transform on a simpler
operator which would differ from $H_c$ by a trace-class operator
\cite{lqc_spectrum2}.

To identify the eigenstates, thus solving the equation $\sum_J \alpha_J\,H_c\,|J\ra\,=\,\beta \sum_J \alpha_J \,|J\ra$, we have to solve the recursion relation:
\be
\forall J\in\N,\quad
2\vphi(J)\alpha_J
+e^{+i\theta}\psi(J)\alpha_{J-1}
+e^{-i\theta}\psi(J+1)\alpha_{J+1}
\,=\,
\beta \alpha_J.
\label{recursioncrit}
\ee
Since this is a second order recursion relation, we expect a two-fold degeneracy for the eigenvectors. Moreover, following the results obtained in the framework of loop quantum cosmology, we expect an asymtptotics of the eigenvectors of the type $\alpha_J \sim \exp(ik\ln J)/\sqrt{J}$ \cite{lqc_spectrum2,merce,lqc_asympt}. This is compatible with the intuition that the frequency $\om$ of oscillations in $J$ vanishes in the critical regime, since we now have oscillations in $\ln J$ which are much slower. Nevertheless, due to the initial relation for $J=0$, the coefficient $\alpha_1$ is entirely determined by the initial coefficient $\alpha_0$. Therefore, the whole sequence $(\alpha_J)_J$ is determined by a single initial condition $\alpha_0$. This kills the two-fold degeneracy and implies that the eigenvalues have a trivial multiplicity of 1.

Following this line of thought, let us insert the following ansatz
in the recursion relation:
\be
\alpha_J\sim \f{(-1)^J}{\sqrt{J}} \,e^{i\theta J}\,e^{ik\ln J},
\ee
with $k\in\R$.
We insert this ansatz in the relation above \Ref{recursioncrit} and we study the large $J$ behavior pushing the approximation to second order in $J$, that is in ${\cal O}(1)$~:
\beq
\vphi(J)&=&J^2+NJ+(N-1),\nn\\
\psi(J)&\sim&
J^2+(N-1)J+\left(\f N2-1\right),\nn\\
\psi(J+1)&\sim&
J^2+(N+1)J+\left(\f {3N}2-1\right),\nn\\
\alpha_{J-1}&\sim&
-e^{-i\theta}\,\alpha_J\,
\left[
1-\left(ik-\f12\right)\f1J+\left(\f38-\f{k^2}2-ik\right)\f1{J^2}
\right]\,,\nn\\
\alpha_{J+1}&\sim&
-e^{+i\theta}\,\alpha_J\,
\left[
1+\left(ik-\f12\right)\f1J+\left(\f38-\f{k^2}2-ik\right)\f1{J^2}
\right]\,.\nn
\eeq
%
%
A straightforward calculation allows to check that this ansatz does
indeed satisfy approximatively (at the order ${\cal O}(1)$) the
recursion relation for an eigenvalue easily expressed in terms of
the frequency $k$~:
\be
\beta=\f14+k^2.
\ee
As expected, this leads to a strictly positive spectrum, with a two-fold degeneracy since $\beta$ is in $k^2$. This gives us a first clue about the spectral properties of $H_c$.

\begin{conj}
Considering the critical regime with $2|\sigma|=\lambda$ with $\lambda>0$, then the (generalized) spectrum of our Hamiltonian $H$ is continuous and covers $]\f\lambda 8,+\infty[$. All the eigenvalues have a multiplicity 1.
\end{conj}

We can check the orthogonality of the eigenvector ansatz~:
\be
\sum_{J}\f1{J}e^{i(k-\tk)\ln J}
\sim
\int_0^{+\infty}dx\,x\,e^{i(k-\tk)\ln x}
\sim
\int_{\R}dy\,e^{i(k-\tk)y}
\propto
\delta(k-\tk).
\ee
Now, in order to get a precise prediction for the asymptotics of the eigenvectors, we need to check how the initial condition collapses the apparent two-fold degeneracy to a trivial multiplicity. A priori, it does not make sense to apply the initial condition relation between $\alpha_0$ and $\alpha_1$ to our asymptotic ansatz for large $J$. Moreover, our ansatz clearly diverges for $J=0$.

\begin{figure}[h]
\begin{center}
\begin{minipage}{50mm}
\reflectbox{\includegraphics[angle=90,height=30mm]{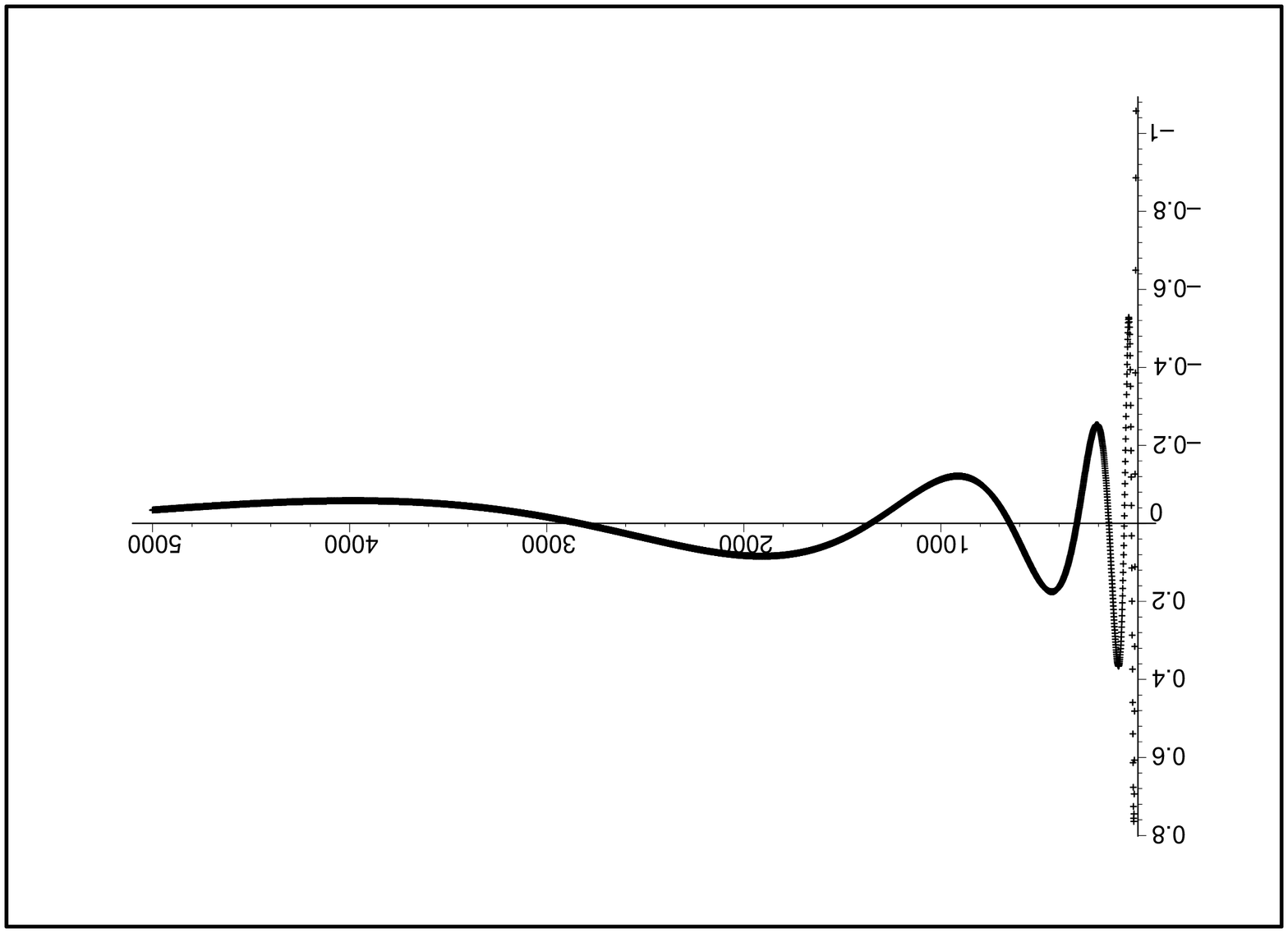}}
\end{minipage}
\begin{minipage}{50mm}
\reflectbox{\includegraphics[angle=90,height=30mm]{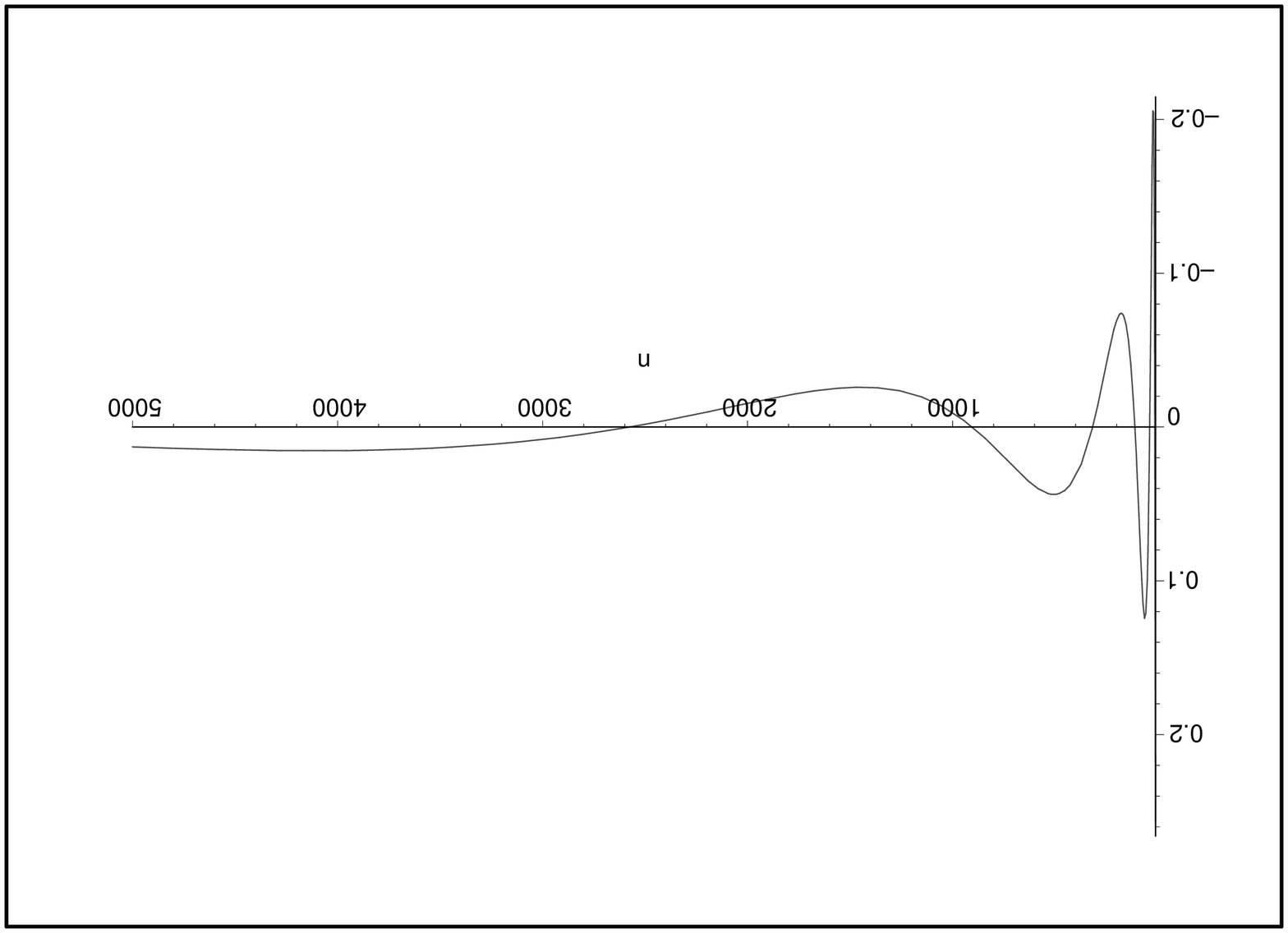}}
\end{minipage}

\begin{minipage}{50mm}
\reflectbox{\includegraphics[angle=90,height=30mm]{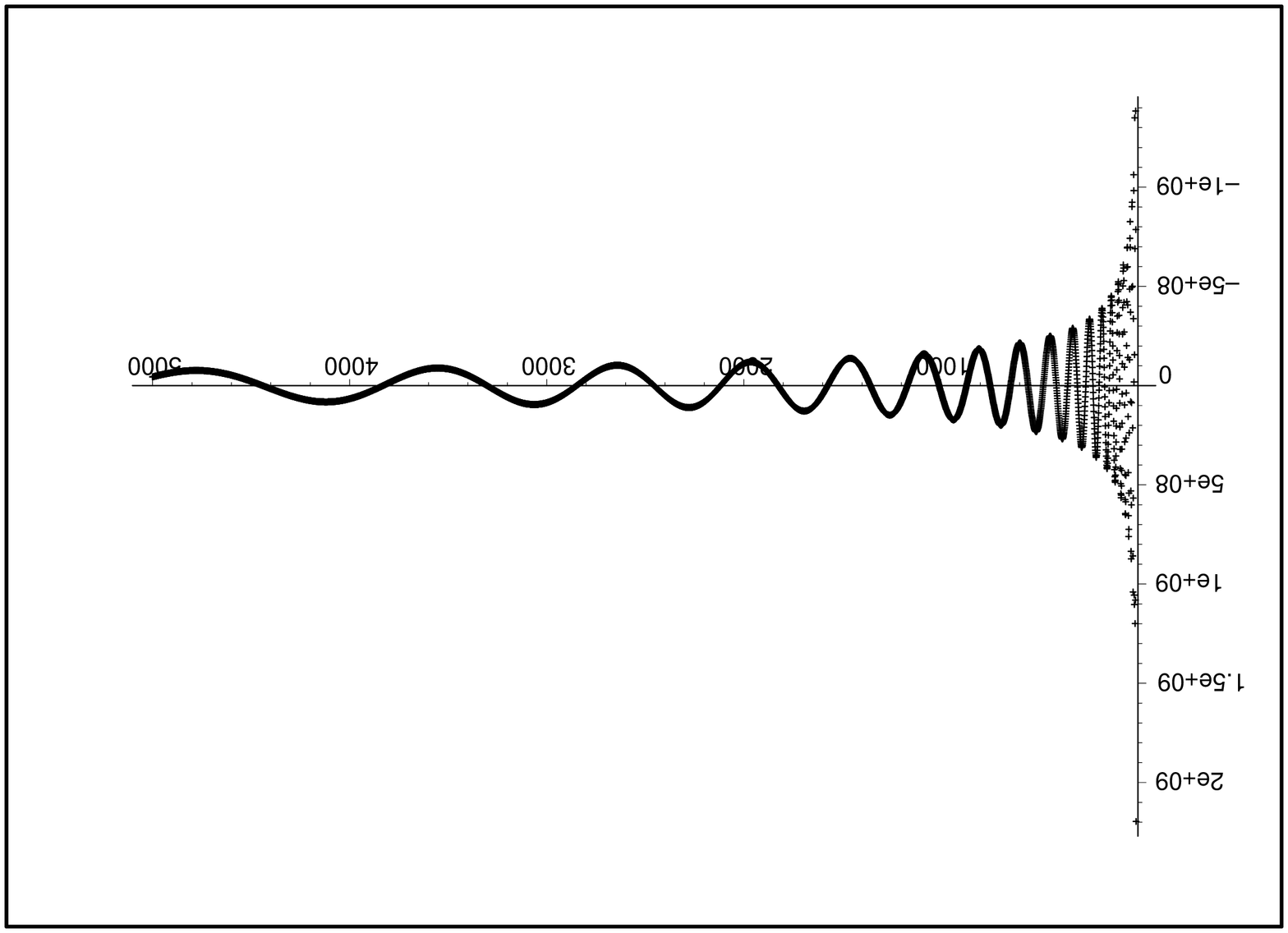}}
\end{minipage}
\begin{minipage}{50mm}
\reflectbox{\includegraphics[angle=90,height=30mm]{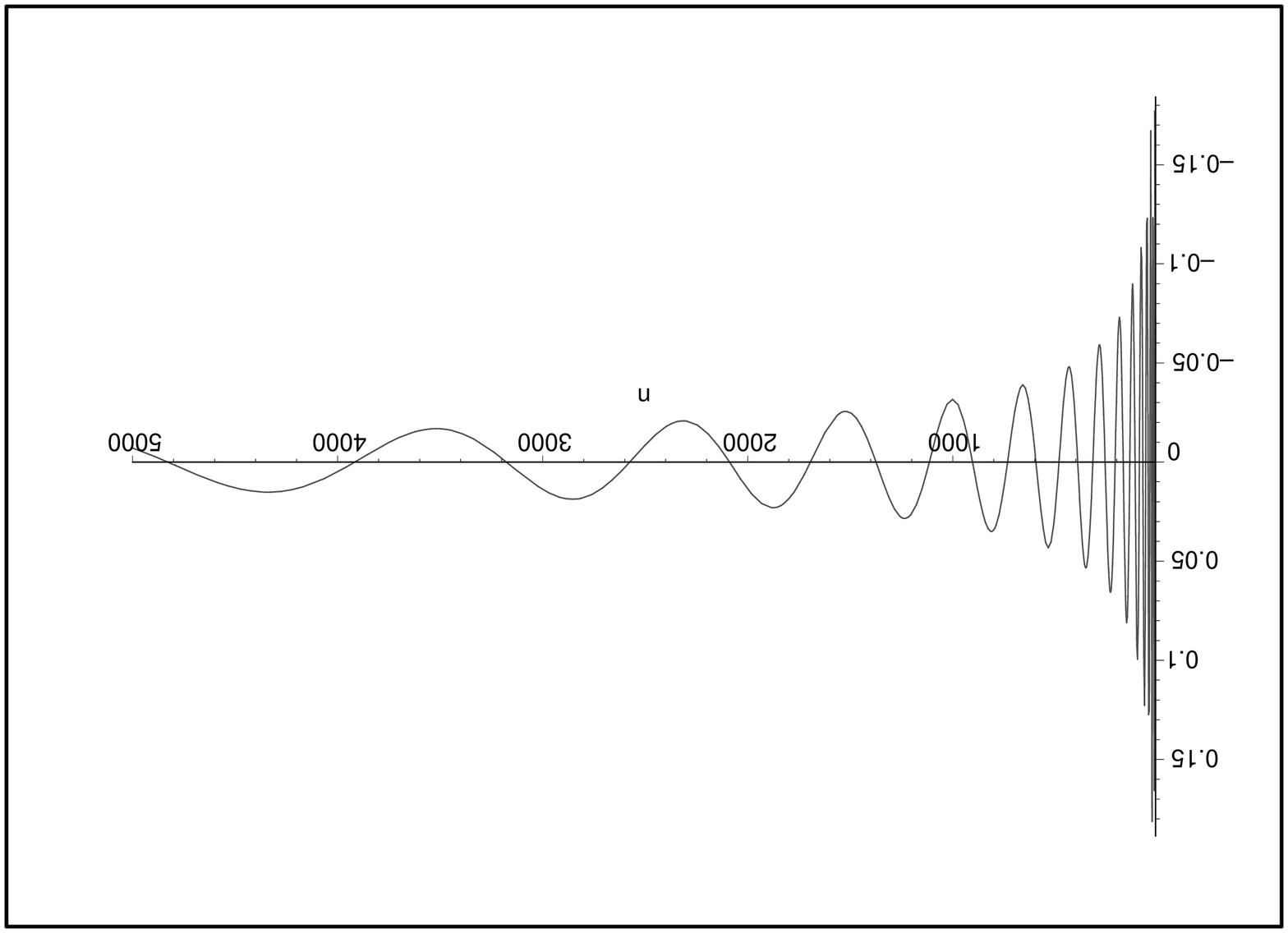}}
\end{minipage}

\caption{Plots in the critical regime for the coefficients $\alpha_J$ for values $N=4$, $\sigma=-1$ and two different frequencies $k=3$ on the top and $k=15$ below. The plots of the left are computed exactly using the recursion relation, while the plots of the right are given by the asymptotic formula. We notice that we have the right oscillatory behavior and the correct overall scaling in $1/\sqrt{J}$, but we still miss approximative predictions for the full amplitude and the initial off-shift of the oscillations.
\label{asymptplots}}
\end{center}
\end{figure}

We set  the initial condition $\alpha_0$ to be real, for instance we choose  $\alpha_0=1$. A first remark is that all the coefficients $\alpha_J$ are then real too up to the $\exp(i \theta J)$ phase due to the coupling $\sigma$ since the coefficients of the recursion relation and the eigenvalue $\beta$ are themselves real. Therefore, a more precise ansatz for the asymptotics of the eigenvectors of large areas $J$ is:
\be
\alpha_J
\,\sim\,
\cA(k)\,\f{(-1)^J}{\sqrt{J}} \,e^{i\theta J}\,\cos(k\ln J+\phi(k)),
\ee
where $\phi(k)$ is a phase depending on the eigenvalue and to be determined, and $\cA(k)$ is the amplitude of the oscillations which depends only on $k$ and is proportional to the initial condition $\alpha_0$.
Unfortunately, we do not yet have any analytical predictions or
simple numerical fits for $\cA(k)$ and $\phi(k)$. Nevertheless, we
provide a couple of plots to illustrate the asymptotical behavior of
the eigenvectors in fig.\ref{asymptplots}.

\medskip

We now turn to the strong coupling case, with $|\lambda|<2|\sigma|$ which correspond to our oscillatory regime. The recursion relation corresponding to the eigenvalue equation, $H\,\sum_J\alpha_J\,|J\ra\,=\,\beta \,\sum_J\alpha_J\,|J\ra$ is as before:
\be
\forall J\in \N,\quad
\lambda\vphi(J)\alpha_J +\bar{\sigma}\psi(J)\alpha_{J-1}+{\sigma}\psi(J+1)\alpha_{J+1}
\,=\,
\beta\alpha_J.
\ee
As pointed out earlier, in this strong coupling, with
$|\lambda|<2|\sigma|$, we can define a real frequency $\om$ in terms
of the couplings:
\be
\lambda+\bar{\sigma} e^{-i\om}+{\sigma} e^{+i\om}=0.
\ee
This allows us to pick up an oscillatory ansatz for a leading order approximation of the eigenvectors:
\be
\alpha_J=\f1 {J+c} e^{i\om J}.
\ee
Inserting this in the recursion relation and pushing the approximation to the next-to-next-to leading order in $J$, that is to the order ${\cal O}(1)$, we use the previous approximations for the factors $\vphi(J)$ and $\psi(J)$, and the following approximations for the coefficients:
$$
\alpha_{J\pm1}\,\sim\, \alpha_J \,e^{\pm i\om}\,
\left(
1\mp\f 1J +\f{1\pm c} {J^2}
\right).
$$
A straightforward calculation gives us the following eigenvalue:
\be
\beta\,=\, \left(\f N 2 -c\right)\,(\bar{\sigma} e^{-i\om}-{\sigma}e^{+i\om}).
\ee
This is a surprising result: $\beta$ is complex for complex values of the parameter $c$, and even purely imaginary when $c\in\R$. Moreover the coefficients $\alpha_J$ are $L^2$ since they go as $1/J$ for large $J$'s. Therefore, we have normalizable eigenstates corresponding to non-real eigenvalues. This leads us to the following statement.

\begin{conj}
Considering the oscillatory regime with $2|\sigma|>|\lambda|$ then the Hamiltonian $H$ is not essentially self-adjoint and its spectrum covers the whole complex plane $\C$.
\end{conj}

This claim is very similar to what happens to the Hamiltonian constraint in loop quantum cosmology for a positive cosmological constant $\Lambda>0$ \cite{KLP}.

\medskip

To conclude, the spectral properties of the Hamiltonian $H$ are well-understood, although we do not have rigorous proofs in the critical and oscillatory regimes and we do not have the exact discrete spectrum in the weak coupling regime. Nevertheless, we know the spectrum of $H$ in all three cases and we have the explicit leading order asymptotics of the eigenvectors in the critical and oscillatory regimes.

However, at the end of the day, it might still be better to work with the $\sl_2$-Hamiltonian $\hh$ whose spectral properties are entirely known and proved. For instance, its discrete spectrum in the weak coupling is explicitly known. Moreover, we know how to exponentiate it into $\SU(1,1)$ group elements, which would define the finite time evolution of the system.

\section{Comparing with Loop Quantum Cosmology}
\label{toLQC}

As already shown in \cite{carlo1,carlo2}, the 2-vertex graph is a perfect setting to derive a quantum cosmology sector from the full loop quantum gravity theory. More precisely, the authors of these papers show how their model was related to loop quantum cosmology. In this section, we similarly underline the explicit relation with the loop quantum cosmology framework.

\subsection{Almost the Same Mathematics}

Comparing our Hamiltonian operator(s) to the loop quantum cosmology framework (e.g. \cite{lqc_asympt}), we immediately notice some substantial differences:
\begin{itemize}
\item The basis states are labeled by a volume variable $v$ instead of our total area variable $J$, and the (modified) holonomy operator induce constant shifts in the volume $v\arr v\pm 4$ which can not be interpreted as constant shifts in the area assuming that the area goes as usual as $J\sim v^{\f23}$.

\item The volume variable is a priori real, $v\in\R$, although superselection rules restrict it to a discrete lattice.

\item The volume $v$ is a priori not bounded by below and is not restricted kinematically to be positive. The restriction to a $v>0$ sector occurs at the dynamical level.

\end{itemize}

From this perspective, our framework is much closer to the original loop quantum cosmology setting initially introduced by Bojowald \cite{bojowald_vol,bojowald_ham}. Despite these differences, the mathematics of the gravitational part of the Hamiltonian constraint in loop quantum cosmology (LQC) are very similar to the mathematics of our Hamiltonian operator.

Indeed, there exist various orderings which have been investigated
in isotropic LQC, for example the APS prescription \cite{aps} or the
sLQC simplified model (``s" standing for solvable) \cite{slqc} or
the MMO ordering \cite{mmo}, but they all have the same leading
order structure at large volume. More precisely, looking at the flat
case with a vanishing cosmological constant $\Lambda=0$, the LQC
framework usually studies two related operators, the gravitational
part of the Hamiltonian constraint $\hat{C}_{gr}$ and the evolution
operator $\hat{\Theta}$, which differ from one another by a density
factor.  Considering the asymptotic regime, $v\gg 1$, the action of
these two operators reads as (e.g. \cite{lqc_asympt}):
\be
\hat{C}_{gr}\,|v\ra
\,\sim\,
(A(v+2)+A(v-2))\,|v\ra-A(v+2)\,|v+4\ra- A(v-2)\,|v-4\ra,
\ee
\be
\hat{\Theta}=\sqrt{v}\hat{C}_{gr}\sqrt{v}
\quad\Rightarrow\quad
\hat{\Theta}\,|v\ra
\,\sim\,
v(A(v+2)+A(v-2))\,|v\ra-\sqrt{v(v+4)}\,A(v+2)\,|v+4\ra-\sqrt{v(v-4)}\, A(v-2)\,|v-4\ra\,
,
\ee
with $A(v)\propto v$ at leading order in the asymptotic limit.
Comparing these expressions to the action of our Hamiltonian operator(s) \Ref{hamiltonian}, it is obvious that they have the same action at leading order if we substitute the LQC variable $v$ by our total area label $J$, or more exactly by setting $v\,\equiv\, 4J$. Indeed the coefficients of $\hat{C}_{gr}$ grow as $v$ and the coefficients of $\hat{\Theta}$ grow as $v^2$~:
\beq
\hat{C}_{gr}\,|v\ra
&\propto&
2v\,|v\ra-v\,|v+4\ra- v\,|v-4\ra\,,\nn\\
\hat{\Theta}\,|v\ra
&\propto&
2v^2\,|v\ra-v^2\,|v+4\ra- v^2\,|v-4\ra\,,
\eeq
up to an overall factor. This leads us to two conclusions. First,
the evolution operator $\hat{\Theta}$ corresponds to our Hamiltonian
operator $H$, whose coefficients grow as $J^2$, while the LQC
Hamiltonian constraint corresponds to our $\sl_2$-Hamiltonian $\hh$,
whose coefficients grow as $J$ at leading order. Therefore, the
spectral properties of the LQC operators and of our Hamiltonian
operators will be very similar and we can apply the techniques
already developed in LQC to our framework
\cite{lqc_spectrum1,lqc_spectrum2,lqc_spectrum3}, and vice-versa.

Second, these LQC operators for the flat case $\Lambda=0$ clearly
correspond to our critical regime with $\sigma=-\lambda/2$. One can
directly compare the properties of the LQC operators to ours in this
regime and we see, for example, that the eigenstates have exactly
the same leading order asymptotics respectively in $\exp(ik\ln
v)\sqrt{v}$ and $\exp(ik\ln J)\sqrt{J}$.

Finally, this identification of the vanishing cosmological constant
case $\Lambda=0$ with our critical regime leads to the natural idea
that the various regimes of our operators correspond to the
different signs of $\Lambda$. We discuss this correspondence in more
details below.

At the end of the day, the present comparison between LQC and our framework is rather superficial. A deeper understanding of the precise relation at the physical level could be achieved when matter or anisotropies/inhomogeneities are included into our 2-vertex model. In particular, this would allow to determine whether our choice of area variable versus the standard LQC choice of a volume variable still leads to sensible physical predictions. This will be investigated in future work \cite{inprep_new}.

\subsection{The Physical Meaning of the Couplings and the Cosmological Constant}
\label{constant}

We would like to look at the way the  cosmological constant is
usually included in isotropic loop quantum cosmology. Following
\cite{KLP}, the cosmological constant simply leads to a diagonal
volume contribution to the Hamiltonian constraint:
\be
\hat{C}_{gr}\,|v\ra
\,=\,
(A(v+2)+A(v-2))\,|v\ra-A(v+2)|v+4\ra-A(v-2)|v-4\ra\,
-\Lambda \hat{V}\,|v\ra\,,
\ee
where we have taken $v>0$ for the sake of simplicity. The volume
operator $\hat{V}$ is obviously assumed to be diagonal:
$$
\hat{V}\,|v\ra\,=\,V_0\, v|v\ra\,,
$$
and the coefficient $A(v)$ is assumed to be linear in $v$ in the
asymptotic limit:
$$
A(v)\,\sim\, 2A_0\,v.
$$
The units $A_0$ and $V_0$ are usually defined in terms of the
gravitational constant $G$, the Planck constant $\hbar$ and the
Immirzi parameter $\gamma$. The interested reader can find the
explicit expressions in \cite{KLP}.
Since the basic shifts in the label $v$ are $\pm 4$, it is natural
to identify $v\equiv 4J$ in order to write explicitly the
correspondence with our $\U(N)$ framework. Moreover, we study
$\hat{\Theta}=\sqrt{v}\hat{C}_{gr}\sqrt{v}$ instead of
$\hat{C}_{gr}$. Translating it in terms of the variable $J$, we get
at leading order~:
\be
\hat{\Theta}\,|J\ra
\,\sim\,
16(4A_0-\Lambda\, V_0)\,J^2 \,|J\ra
-32 A_0(J+\f12)\sqrt{J(J+1)}\,|J+1\ra
-32 A_0(J-\f12)\sqrt{J(J-1)}\,|J-1\ra.
\ee
We compare it directly with the expression of our Hamiltonian operator $H$ given in \Ref{hamiltonian}. The two expressions match exactly at leading order in $J$. It is now clear that when $\Lambda=0$, we are in the critical regime. On the other hand, as soon as $\Lambda\ne 0$, we leave the critical regime and enter either the discrete or oscillatory regimes.
More precisely, we have the following correspondence between our couplings and the LQC parameters:
\be
\lambda\,\equiv\,16(4A_0-\Lambda V_0),\qquad
\sigma=\bar{\sigma}\,\equiv\,-32 A_0.
\ee
Thus when $\Lambda=0$, we have $\sigma =-\lambda/2$ and we are in the critical regime. When $\Lambda>0$, but still close to $0$, we have $0<\lambda<2|\sigma|$ and we are in the strong coupling regime with a non-essentially self-adjoint Hamiltonian operator $H$. Finally, when $\Lambda<0$, we have $\lambda>2|\sigma|$ and we are in the weak coupling regime with a strictly positive discrete spectrum of the Hamiltonian $H$. These 3 cases perfectly fit with the results obtained in LQC \cite{KLP,paw_cosmo}.

This correspondence is however very rough. It needs to be refined on at least two points:
\begin{itemize}

\item Here we made the substitution $v\arr 4J$ at the purely mathematical level. Of course, physically, we can not replace the volume variable $v$ by the area label $J$. Thus we would need to understand better the role and the validity of such a substitution and in particularly we should study the expectation value of the volume of our isotropic states $|J\ra$.

\item We need to include matter in our 2-vertex model to truly compare it with the full Hamiltonian constraint and the exact LQC results.

\end{itemize}

Besides these set-backs, it is clear that the cosmological constant contributes a diagonal term to the Hamiltonian in the $|J\ra$ basis (since the value of the volume should be given uniquely by the total area $J$ for isotropic configurations) so that shifting $\Lambda$ is naturally reflected in the relative values of the couplings $\lambda$ and $\sigma$. From there, it is natural to identify the three regimes of our Hamiltonian, discrete or critical or oscillatory, respectively to the three sectors $\Lambda<=>0$.

From this perspective, we have a physical interpretation to arbitrary real  values of the coupling $\sigma$. We would still need to understand the physical meaning of the possible complex phase of that coupling.

\subsection{Going further with LQC?}

So far, we have shown how our framework resembles the LQC
framework and how the LQC techniques  can be used to study the
properties of our Hamiltonian. Moreover, the LQC framework is much
more developed than ours and includes matter fields and investigates
various cosmological settings. Thus, a natural question is: what can
our approach bring to LQC? We would like to present two points on
which our approach can bring something new to LQC.

Our first point is technical. It's the remark that our operators
acting on the homogeneous subspace form a $\sl_2$ algebra.  This
allows us to characterize our homogeneous sector as an irreducible
unitary representation of $\SL(2,\R)$ and to construct a Hamiltonian
operator $\hh$ as a $\sl_2$-Lie algebra element whose spectrum and
eigenvectors are explicitly known (in all cases of the couplings).
Since the structure of the actions of our Hamiltonian operator(s)
and of the LQC Hamiltonian constraint are very similar, we can
expect that a particular choice of ordering in LQC would also lead
to a $\sl_2$-structure. This would allow for a more refined
algebraic characterization of the LQC Hilbert space and of the
spectrum of the Hamiltonian of isotropic LQC.

Our second point has deeper consequences. It's on the crucial role of $\U(N)$ as the symmetry at the quantum level reducing our full Hilbert space of spin network states to the subspace of isotropic and homogeneous states. Indeed, the standard loop quantum cosmology framework relies on the implementation of the requirement of isotropy/homogeneity at the classical level: we reduce the degrees of freedom at the classical level, and then quantize the reduced phase space using loop quantum gravity techniques. Here the procedure is done the other way round: we start with the full loop quantum gravity and reduce the degrees of freedom directly at the quantum level by first selecting the underlying 2-vertex graph with an arbitrary number $N$ of edges and then imposing the invariance under $\U(N)$ in order to reduce our Hilbert space to isotropic and homogeneous states. From this viewpoint, the $\U(N)$ symmetry could be the tool, missing up to now, necessary  to perform the symmetry reduction directly at the quantum level from loop quantum gravity to isotropic cosmological situations.

\section{Going Beyond the 2-Vertex Model}

In this section we will comment on the main limitations of our model and
some possible generalizations of it. More precisely, we will
propose a tentative model with 3+$N$ vertices where, in principle, it
could be possible to study certain additional physically relevant processes.

Let us start by sketching, in first place, some possible generalizations of
our model (in order of increasing difficulties):

\begin{itemize}

\item There are neither inhomogeneities nor anisotropies in our model since we have restricted ourselves to work on the $\U(N)$-invariant space,
which restricts to isotropic states ($|J\rangle$) only labeled by
the total boundary area $J$. We could relax this requirement
and study the dynamics of non-$\U(N)$-invariant states. Since our
dynamics is $\U(N)$-invariant, irreducible representations of the
$\U(N)$ group would still be stable under the time evolution. It
would be interesting to check what type of anisotropies and inhomogeneities
correspond to each of these irreducible $\U(N)$ representations in
the spin network space of our 2-vertex graph, use $\U(N)$ to
classify anisotropies and compare to the standard expansion in
terms of angular momentum.

\item We could go further and break our requirement of $\U(N)$ invariance
of the dynamics.  We could do this by hand, by choosing operators
which are not invariant, but it does not seem physically motivated
to break a symmetry for free with no clear purpose.  Moreover, the
dynamics of the two vertices is exactly the same due to our choice
of dynamical operators, which are symmetric, so there is no freedom
for local fluctuations.  Furthermore, there is no matter in our
simple model.  We could add a scalar field on the 2-vertex graph
(like in \cite{carlo1}). Matter could provide us with a source to
break the $\U(N)$ invariance and introduce local fluctuations in the
model. It would also allow to obtain a true (though still very
simple) cosmological model and compare it with the results in loop
quantum cosmology.

\item Our two vertices are strongly glued with the matching conditions.
They grow, shrink and rotate, all together with no freedom for
different evolutions for the two vertices. We cannot relax this
without changing the underlying graph and using a more complicated
spin network setting. This would allow us to represent physical
processes such as the rotation of one vertex with respect to another
or the evaporation of a (horizon) boundary. Here we propose to
consider the next simplest graph, a 3+$N$ vertex graph, which allows
to decouple the two original vertices and recover the independent
$\U(N)$-actions on the intertwiner spaces of both vertices.

\end{itemize}

We postpone these points to future investigation. Nevertheless, we
would like to describe here the 3+$N$ vertex graph that we just
mentioned. It is the simplest graph beyond the 2-vertex graph and we
will sketch how it would allow a richer physical arena in which more
processes could take place.

\medskip

The main physical limitations of the 2-vertex graph considered so
far are direct consequences of the gluing between both vertices. We
have seen how this model is good enough to describe homogeneous
cosmological situations, and it would be also possible to consider
the introduction of inhomogeneities by relaxing the $\U(N)$
symmetry, as just commented above. However, an interesting point of
view is to regard the two vertices as representing an ``in'' and an
``out'' regions. Adopting this point of view, we would like to be
able to describe physical situations in which the ``in'' region has
some kind of dynamical evolution with respect to the ``out'' region.
Examples of these processes could be evaporation of the ``in''
region towards the exterior, or rotations of the interior. The present 2-vertex model
turns out to be too simple to describe this kind of physical
situations. For instance, in particular, the ``out'' region cannot
exist without the ``in'' region, so in an evaporation process (e.g. as in the case of a physical black hole)
causing the ``in'' region to shrink and eventually disappear, the
``out'' region would also be forced to disappear. These limitations
are imposed by the graph itself. Thus in order to introduce enough
freedom to study this kind of processes, we propose to consider a
slightly more sophisticated graph.

\begin{figure}[h]
\begin{center}
\includegraphics[height=35mm]{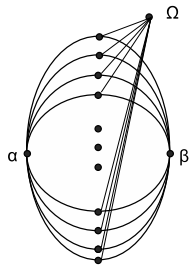}
\caption{The 3-vertex graph: the three vertices $\alpha$, $\beta$,
and $\Omega$ are linked by $N$ edges and we have $N$ additional
(auxiliary) trivalent vertices. \label{3vertex}}
\end{center}
\end{figure}

This generalization  consists in adding a third vertex to
the graph, which will be part of the ``out'' region. This new vertex
$\Omega$ is linked to the previous graph by $N$ edges, each one
meeting one of the old edges at a new trivalent vertex $a_i$. The
resulting picture is shown in figure \ref{3vertex}. The interior
region continues to be the vertex $\alpha$ and the boundary is the
surface surrounding it. The $N$ trivalent vertices are considered as part of
the exterior region, but can also be interpreted as representing the physical boundary between the ``in" and ``out" regions.

We will consider operators acting on the new vertices in analogy to
the ones we already have for the simpler 2-vertex model. Thus, we have
operators  $E^{(\Omega)}_{ij}$ and $F^{(\Omega)}_{ij}$ acting on the
vertex $\Omega$ as well as $E^{(a_i)}_{jk}$ and $F^{(a_i)}_{jk}$, with
$i=1,\ldots,N$ and $j,k=\alpha,\beta,\Omega$, acting on the $N$
trivalent vertices.

Now, thanks to this new vertex $\Omega$ it is possible to shrink the
area surrounding the ``in''  region without resulting in a shrinking
of the ``out''  region too. One can even think of the ``in'' region totally
disappearing, and the ``out'' region would still make sense.
Thus, in this framework it is possible to introduce radiation-like
operators. As a first attempt we propose an operator of the kind

\be
R_{ij}=F^{(\alpha)}_{ij}E^{(a_i)}_{\Omega\alpha}F\dag{}^{(\Omega)}_{ij}
E^{(a_j)}_{\Omega\alpha}\,.
\ee

The action of this operator over the graph is to decrease the spin
associated to the edges $i$ and $j$ of the interior vertex, and to
increase the spins of the same edges of the vertex $\Omega$. In this
sense, the action of the operator decreases the area of the boundary
by ``radiating'' some spin units to the exterior region. The vertex
$\Omega$ could be interpreted in this framework as carrying the
radiated field.

\begin{figure}[h]
\begin{center}
\includegraphics[height=20mm]{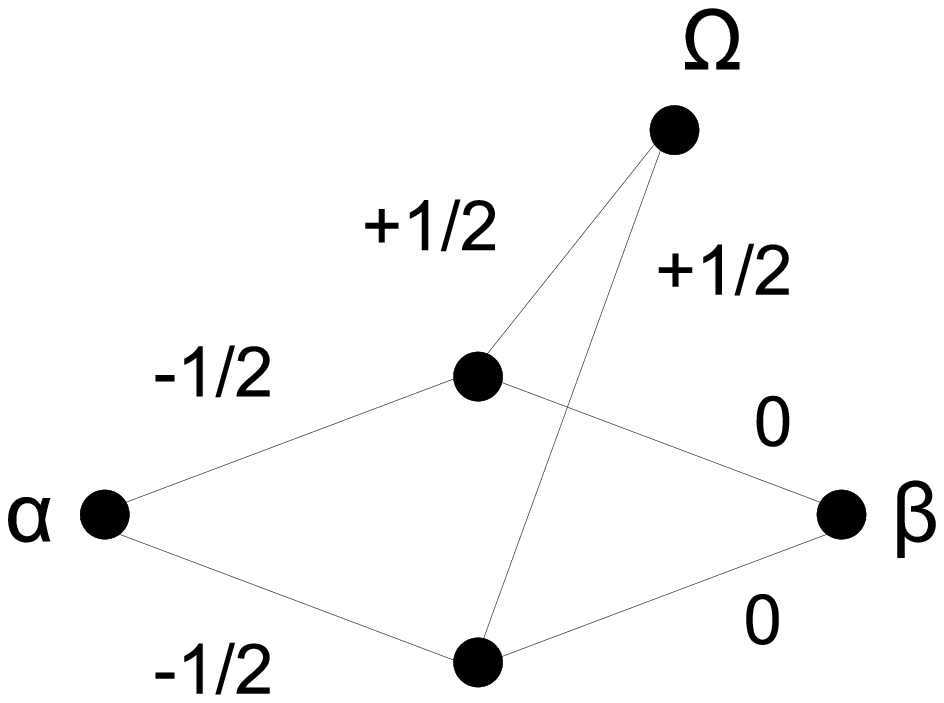}
\hspace{20mm}
\includegraphics[height=20mm]{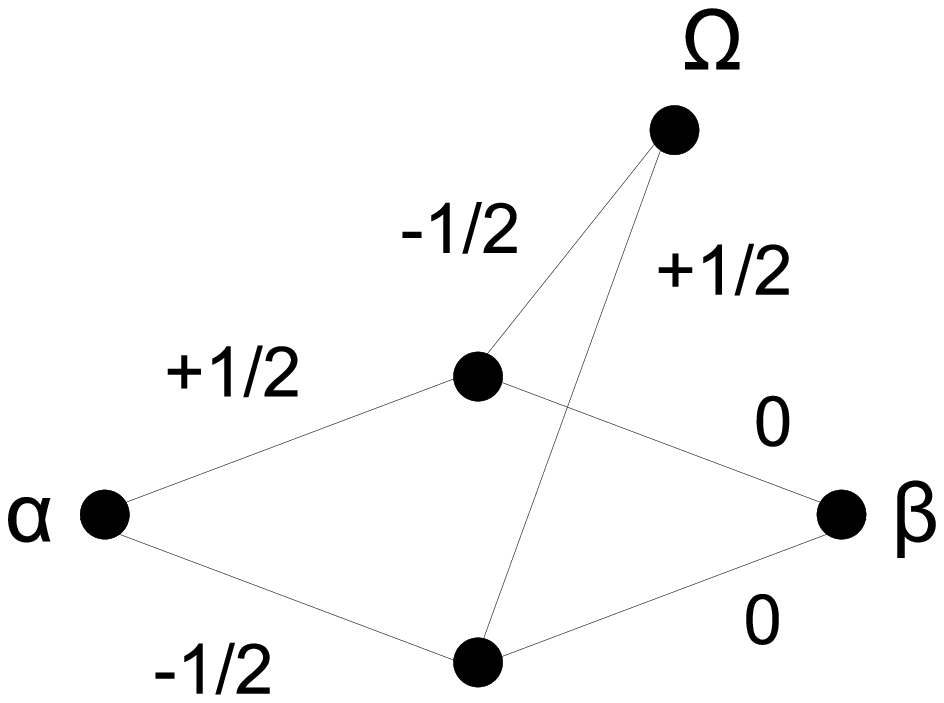}
\end{center}
\caption{The action of the radiation operator on the left and the rotation-like operator on the right: we shift the spins on the legs $i,j$ around the vertices $\alpha$ and $\Omega$ without affecting the spins living on the legs around the vertex $\beta$.\label{3vertexops}}
\end{figure}

In a similar way, one can introduce operators generating
rotation-like transformations, which would change the
representations living on the $N$ legs of the vertex $\alpha$ but
without changing the total area of the boundary around $\alpha$ and
without affecting the representations living on the $N$ legs of the
vertex $\beta$.
It will be necessary to check whether the commutators of these radiation operators and rotation-like operators satisfy a closed algebra.
%
We postpone the study of the details of these
generalizations, as well as the implementation of the dynamics for
this new 3+$N$ model, to future work \cite{inprep}.

%
%

\section{Conclusions and Outlook}

%
%
%
%
%

The $\U(N)$ framework recently introduced in \cite{un1,un2,un3}
represents a new and refreshing way to tackle the issue of
dynamics within the loop quantum gravity framework.  Our motivation is
not to solve the full set of constraints but to propose a Hamiltonian operator
which describes the dynamical behavior of a given system, following
the usual procedure even in general relativity.  Here, we have
presented a simple {\it 2-vertex model} --based literally on graphs with 2 vertices and an arbitrary number $N$ of edges-- and we have proposed a tentative Hamiltonian for the dynamics of this specific system.

In those previous papers \cite{un1,un2,un3}, it was shown that the
space of $\SU(2)$ intertwiners for loop quantum gravity carries a
natural action of the unitary group $\U(N)$~:  the Hilbert space of
intertwiners for a given number $N$ of edges and with a fixed total
area $J$ is identified as an irreducible $\U(N)$ representation.
This construction is based on the Schwinger representation of
$\SU(2)$ in terms of harmonic oscillators. Using this tool,  a set
of operators invariant under global $\SU(2)$ transformations have
been identified~: the $\u(N)$ generators $E_{ij}$ and
annihilation/creation operators $F_{ij},F\dag_{ij}$, which endow the
full space of $N$-valent intertwiners with a Fock space structure.
%
%

In this work we have described a very simple model for loop quantum
gravity, based on the 2-vertex graph described above.  We have the space of intertwiners for each
vertex, plus some matching conditions coupling the two intertwiners
by imposing that each edge $i$ of the graph carries a unique spin
label $j_i$ as seen from both vertices (these conditions reflect
that one edge carries a single $SU(2)$ irreducible representation).
Taking this into account, we have introduced operators
$e_{ij},f_{ij}, f\dag_{ij}$ that commute with the matching
conditions and constitute the building blocks for the dynamics on
our 2-vertex graph. Going further, we have identified these matching
conditions as  the Cartan subalgebra of a larger $\u(N)$ algebra
acting on both vertices, and we have shown that the invariance under
this global $\U(N)$ symmetry implies the restriction to isotropic
homogeneous states $|J\ra$ which are not sensitive to
area-preserving deformations of the boundary between $\alpha$ and
$\beta$. This provides our $\U(N)$ symmetry with a physical interpretation.
This is actually one of the key results presented here.
Indeed it could be viewed as a key step towards a full understanding
between the general loop quantum gravity (LQG) framework and the
symmetry-reduced loop quantum cosmology (LQC). More precisely, the
standard framework of loop quantum cosmology is based on a symmetry
reduction at the classical level followed by a quantization using
loop gravity techniques. Since the symmetry reduction is done before
the quantization, we have lost the connection between the Hilbert
spaces of LQG and LQC and we don't know how to perform the symmetry
reduction directly at the level of the full quantum theory. A first
hint towards solving this issue was provided by the approach of
Rovelli-Vidotto \cite{carlo1} which is also based on the 2-vertex
graph with $N=4$ edges. They propose to consider the total area
$J\equiv\sum_i j_i$ as the homogeneous degree of freedom and to
interpret the individual fluctuations of the spins $j_i$ on each
edge as the anisotropy/inhomogeneous degrees of freedom. They
implemented this separation of variables by hand when studying the
dynamics of their system. Here we went one step further and
identified explicitly the symmetry which selects the homogeneous
states. Now the reduction to the Hilbert space of isotropic
homogeneous states can be achieved through a simple $\U(N)$ group
averaging. We hope that this technique will be relevant to
understanding how to derive quantum cosmology  from the LQG
framework directly at the quantum level. This will be investigated
in more details in future work \cite{inprep_new}.
%

Another interesting point that we would like to stress is that we have
studied the relation between our operators and the usual holonomy
and flux (derivation/grasping) operators standardly used in LQG. This provides an explicit bridge between our formulation and the standard way to formulate dynamics in LQG. From this point of view, we hope that our formalism will be interesting to study the general issue of the LQG dynamics on arbitrary graphs, more complicated than the 2-vertex graph used in the present work.
%

The model that we have developed here has a simple enough algebraic structure to allow us to
construct a Hamiltonian that respects isotropy and evolves homogeneous states into homogeneous states. In order to be able to build such a Hamiltonian, we have defined
$\U(N)$-invariant operators $e$ and $f$ as linear combinations
(summing over all the edges in order to have homogeneity) of
our building block operators $e_{ij}$ and $f_{ij}$. It turns out that these invariant operators simply satisfy a  $\sl_2$-algebra. Using the operators $f$
and $f\dag$ as (area) annihilation and creation operators, we have
studied their action on an homogeneous state and we have discussed the interpretation of  the creation operators $(f\dag)^J$ as ``black hole'' creation operators.
This allowed us to define a $\U(N)$-invariant Hamiltonian, which is unique up to a renormalization by a factor depending on the total boundary area. Here, we have focused on two precise choices of normalization, which allow a direct comparison with the Hamiltonian constraints used in LQC, and we have studied their spectral properties. We hope to be able to apply in the future this analysis of their properties to further develop the derivation of LQC from our 2-vertex LQG model \cite{inprep_new}, following the path opened by Rovelli and Vidotto \cite{carlo1,carlo2}.
%

To conclude, although we have been dealing with a simplified model,
and one has to be cautious in order to extrapolate the results to
more general frameworks, in our opinion, we have obtained sensible
results and it seems that the connections of our 2-vertex quantum
model with LQC and LQG  are fully justified. From this perspective,
we believe that it is plausible to use this kind of simple models as
a testbed for certain dynamical problems within the LQG and LQC
theories. A possible extension of the work would be the
implementation of anisotropies/inhomogeneities and the study of the
stability of homogeneous states under the evolution given by our
Hamiltonian. Finally, we also discussed the possibility of working
on a more complex graph and switching to a $3+N$-vertex graph, which
would allow to construct operators representing physical rotation
and evaporation and thus to simulate the dynamics of physical
quantum black holes in LQG.

\section*{Acknowledgments}

We are specially grateful to Mercedes Mart\'in-Benito for many
discussions on the present work and especially many explanations on
the general framework of loop quantum cosmology and its technical
features.

This work was in part supported by the Spanish MICINN research grants
FIS2008-01980, ESP2007-66542-C04-01 and FIS2009-11893. JD is supported
by the NSF grant PHY0854743, The George A. and Margaret M.
Downsbrough Endowment and the Eberly research funds of Penn State.
IG is supported by the Department of Education of the Basque
Government under the ``Formaci\'{o}n de Investigadores'' program. EL is
partially supported by the ANR ``Programme Blanc" grants LQG-09.
%

\appendix

\section{Diagonalizing the renormalized Hamiltonian}

In this appendix, we study the action of the renormalized
Hamiltonian operator $h$ (with $\sigma\in\R$), whose action on the
homogeneous space $\cHi$ is given by~:
$$
h|J\ra\,=\,
\lambda|J\ra \,+ \sigma|J-1\ra+\sigma|J+1\ra,\qquad
h|0\ra\,=\,
|0\ra \,+\sigma|1\ra.
$$
Looking for eigenvectors $h\,\sum_J \alpha_J\,|J\ra\,=\,\beta\,\sum_J \alpha_J\,|J\ra$, we obtain a simple second order recursion relation on the coefficients $\alpha_J$ of the corresponding state:
\beq
&&\lambda\alpha_0+\sigma\alpha_1=\beta\alpha_0\,,
\label{hinitial}\\
&&\lambda\alpha_J+\sigma\alpha_{J-1}+\sigma\alpha_{J+1}=\beta\alpha_J\,,\quad \forall J\ge 1\,.
\label{hrecursion}
\eeq
Putting aside the initial condition equation \Ref{hinitial} and focusing on the recursion relation \Ref{hrecursion}, we see that the plane waves $\alpha_J=\exp(i\om J)$ with arbitrary frequency satisfy this second order recursion~:
\be
\lambda e^{i\om J}+\sigma e^{i\om (J-1)}+\sigma e^{i\om (J-1)}
\,=\,
(\lambda+2\sigma\cos \om)\,e^{i\om J}.
\ee
The eigenvalues $\beta=(\lambda+2\sigma\cos \om)$ are obviously bounded and the spectrum is doubly degenerate since the two conjugate plane waves $\exp(i\pm\om J)$ correspond to the same eigenvalue. Now, we need to take into account the initial condition  \Ref{hinitial} and this leads to a non-degenerate spectrum with a single plane wave for each eigenvalue/frequency. Indeed, considering the ansatz $\alpha_J=a\cos\om J+b\sin\om J$, the initial condition equation reads~:
\be
\lambda a + \sigma (a\cos\om+b\sin\om)
\,=\,
\beta a
\,=\,
(\lambda+2\sigma\cos \om)\,a.
\ee
This relates the two coefficients $a$ and $b$ of the plane wave,
\be
b\sin\om=a\cos\om,
\ee
which leads to a unique eigenvector for each frequency $\om$:
\beq
&h\,|\om\ra\,=\,(\lambda+2\sigma\cos \om)\,|\om\ra,
\quad
\textrm{with}\quad
|\om\ra&\equiv\, \sum_J (\cos\om\sin\om J +\sin\om \cos\om J)\,|J\ra \\
&&\,=\,
\sum_{J\in\N} \sin\left[(J+1)\om\right] \,|J\ra
\,=\,
\sin\om\sum_J \chi_{\f J 2}(\om) \,|J\ra. \nn
\eeq
Due to the invariance under parity $\om\arr -\om$, we restrict ourselves to positive frequencies $\om\in[0,\pi]$.  Then it is straightforward to check that these eigenstates are orthogonal to each other:
\be
\la \tom |\om\ra
\,=\,
\sum_{J\in \N}\sin\left[(J+1)\om\right]\sin\left[(J+1)\tom\right]
\,=\,
\sum_{n\in \Z}\sin n\om\,\sin n\tom
\,=\,
\f\pi 2\,\delta(\om-\tom).
\ee



\begin{thebibliography}{99}

\bibitem{thomas1}
T. Thiemann,
{\it Quantum Spin Dynamics (QSD)},
Class.Quant.Grav. 15 (1998) 839-873 [arxiv:gr-qc/9606089]


\bibitem{thomas2}
T. Thiemann,
{\it Quantum Spin Dynamics (QSD) II},
Class.Quant.Grav. 15 (1998) 875-905 [arxiv:gr-qc/9606090]

\bibitem{tina1}
K. Giesel and T. Thiemann,
{\it Algebraic Quantum Gravity (AQG) I. Conceptual Setup },
Class.Quant.Grav.24 (2007) 2465-2498 [arXiv:gr-qc/0607099]

\bibitem{spinfoam1}
M.P. Reisenberger and C. Rovelli,
{\it ``Sum over Surfaces'' form of Loop Quantum Gravity},
Phys.Rev. D56 (1997) 3490-3508 [arXiv:gr-qc/9612035]

\bibitem{spinfoam2}
J. Engle, E.R. Livine, R. Pereira and C. Rovelli,
{\it LQG vertex with finite Immirzi parameter},
Nucl.Phys.B799 (2008) 136-149 [arXiv:0711.0146]


\bibitem{carlo1}
C. Rovelli and F. Vidotto,
{\it Stepping out of Homogeneity in Loop Quantum Cosmology},
Class.Quant.Grav.25 (2008) 225024 [arXiv:0805.4585]

\bibitem{carlo2}
M.V. Battisti, A. Marciano and C. Rovelli,
{\it Triangulated Loop Quantum Cosmology: Bianchi IX and inhomogenous perturbations},
arXiv:0911.2653

\bibitem{carlo3}
E. Bianchi, C. Rovelli and F. Vidotto, {\it Towards Spinfoam
Cosmology}, arXiv:1003.3483

\bibitem{un1}
F. Girelli and E.R. Livine,
{\it Reconstructing Quantum Geometry from Quantum Information: Spin Networks as Harmonic Oscillators},
Class.Quant.Grav. 22 (2005) 3295-3314 [arXiv:gr-qc/0501075]

\bibitem{un2}
L. Freidel and E.R. Livine,
{\it The Fine Structure of $\SU(2)$ Intertwiners from $\U(N)$ Representations},
to appear in JMP 2010 [arXiv:0911.3553]

\bibitem{un3}
L. Freidel and E.R. Livine,
{\it $\U(N)$ coherent states for Loop Quantum Gravity},
arXiv:1005.2090

\bibitem{matrixmodel}
E. Borja, J. D\'{\i}az-Polo, I. Garay and E.R. Livine, {\it Matrix Models
for LQG Intertwiners}, in preparation



\bibitem{su11}
L. Freidel, E.R. Livine and C. Rovelli,
{\it Spectra of Length and Area in 2+1 Lorentzian Loop Quantum Gravity},
Class.Quant.Grav. 20 (2003) 1463-1478 [arXiv:gr-qc/0212077]


\bibitem{bh1}
E.R. Livine and D. Terno,
{\it Quantum Black Holes: Entropy and Entanglement on the Horizon},
Nucl.Phys.B741 (2006) 131-161 [arXiv:gr-qc/0508085]

\bibitem{bh1bis}
E.R. Livine and D. Terno,
{\it Bulk Entropy in Loop Quantum Gravity},
Nucl.Phys.B794 (2008) 138-153 [arXiv:0706.0985]

\bibitem{bh2}
J. Engle, K. Noui and A. Perez,
{\it Black hole entropy and SU(2) Chern-Simons theory},
arXiv:0905.3168

\bibitem{bh3}
K. Krasnov and C. Rovelli,
{\it Black holes in full quantum gravity},
Class.Quant.Grav.26 (2009) 245009 [arXiv:0905.4916]

\bibitem{bh4}
I. Agullo, J.F. Barbero G., E.F. Borja, J. D\'{\i}az-Polo and E.J.S.
Villase\~{n}or {\it The combinatorics of the SU(2) black hole entropy in
loop quantum gravity}, Phys.Rev.D80 (2009) 084006 [arXiv:0906.4529]

\bibitem{KLP}
W. Kaminski, J. Lewandowski and T. Pawlowski,
{\it Physical time and other conceptual issues of QG on the example of LQC},
Class.Quant.Grav.26 (2009) 035012 [arXiv:0809.2590]

\bibitem{lqc_spectrum3}
A. Ashtekar, T. Pawlowski, P. Singh and K. Vandersloot, {\it Loop
quantum cosmology of k=1 FRW models}, Phys.Rev.D75 (2007) 024035
[arXiv:gr-qc/0612104]

\bibitem{SG}
K. Schulten and R. Gordon,
{\it Exact Recursive Evaluation of 3J and 6J Coefficients for Quantum Mechanical Coupling of Angular Momenta}, J.Math.Phys.16 (1975) 1961-1970

\bibitem{maite}
M. Dupuis and E.R. Livine,
{\it The 6j-symbol: Recursion, Correlations and Asymptotics},
Class.Quant.Grav.27 (2010) 135003 [arXiv:0910.2425]




\bibitem{merce}
M. Martin-Benito, G.A. Mena Marug\'{a}n and T. Pawlowski, {\it Physical
evolution in Loop Quantum Cosmology: The example of vacuum Bianchi
I}, Phys.Rev.D80 (2009) 084038 [arXiv:0906.3751]

\bibitem{su11davids}
S. Davids, {\it A State Sum Model for (2+1) Lorentzian Quantum
Gravity}, PhD Thesis, Nottingham University, arXiv:gr-qc/0110114

\bibitem{lqc_spectrum1}
{\L}. Szulc, W. Kaminski and J. Lewandowski, {\it Closed FRW model
in Loop Quantum Cosmology}, Class.Quant.Grav.24 (2007) 2621-2636
[arXiv:gr-qc/0612101]


\bibitem{lqc_spectrum2}
W. Kaminski and J. Lewandowski,
{\it The flat FRW model in LQC: the self-adjointness},
Class.Quant.Grav.25 (2008) 035001 [arXiv:0709.3120]



\bibitem{lqc_asympt}
W. Kaminski and T. Pawlowski,
{\it Cosmic recall and the scattering picture of Loop Quantum Cosmology},
Phys.Rev.D81 (2010) 084027 [arXiv:1001.2663]

\bibitem{bojowald_vol}
M. Bojowald, {\it Loop Quantum Cosmology II: Volume Operators},
Class.Quant.Grav. 17 (2000) 1509-1526 [arXiv:gr-qc/9910104]

\bibitem{bojowald_ham}
M. Bojowald,
{\it Absence of Singularity in Loop Quantum Cosmology},
Phys.Rev.Lett.86 (2001) 5227-5230 [arxiv:gr-qc/0102069]

\bibitem{aps}
A. Ashtekar, T. Pawlowski, and P. Singh,
{\it Quantum Nature of the Big Bang: Improved dynamics},
Phys. Rev. D73 (2006) 124038  [arXiv:gr-qc/0607039]

\bibitem{slqc}
A. Ashtekar, A. Corichi and P. Singh ,
{\it Robustness of key features of loop quantum cosmology},
Phys.Rev.D77 (2008) 024046 [arXiv:0710.3565]


\bibitem{mmo}
M. Mart\'in-Benito, G.A. Mena Marug\'an and T. Pawlowski,
{\it Loop Quantization of Vacuum Bianchi I Cosmology},
Phys. Rev. D78 (2008) 064008  [arXiv:0804.3157];\\
M. Mart\'in-Benito, G.A. Mena Marug\'an and J. Olmedo,
{\it Further Improvements in the Understanding of Isotropic Loop Quantum Cosmology},
Phys. Rev. D80 (2009) 104015  [arXiv:0909.2829]



\bibitem{paw_cosmo}
E. Bentivegna and T. Pawlowski,
{\it Anti-deSitter universe dynamics in LQC},
Phys.Rev.D77 (2008) 124025 [arXiv:0803.4446]


\bibitem{inprep}
E.F. Borja, J. D\'{\i}az-Polo, I. Garay and E.R. Livine, {\it Quantum
gravity dynamics on the 3+N vertex graph and black hole
evaporation}, in preparation

\bibitem{inprep_new}
E.F. Borja, J. D\'{\i}az-Polo, I. Garay and E.R. Livine,
{\it Loop Quantum Cosmology from the $\U(N)$
Framework}, in preparation


\end{thebibliography}
\end{document}